\documentclass[twocolumn]{aastex63}
\usepackage{graphicx}
\usepackage{natbib}
\usepackage{placeins}
\usepackage{amsmath}
\usepackage{tgcursor}
\usepackage{booktabs}
\usepackage{enumitem}
\setcitestyle{notesep={ }}

\begin{document}

\title{Molecules with ALMA at Planet-forming Scales (MAPS) XI:\\ CN and HCN as Tracers of Photochemistry in Disks}

\correspondingauthor{Jennifer Bergner}
\email{jbergner@uchicago.edu}

\author[0000-0002-8716-0482]{Jennifer B. Bergner}
\altaffiliation{NASA Hubble Fellowship Program Sagan Fellow}
\affiliation{University of Chicago Department of the Geophysical Sciences, Chicago, IL 60637, USA}

\author[0000-0001-8798-1347]{Karin I. \"Oberg} 
\affiliation{Center for Astrophysics \textbar\ Harvard \& Smithsonian, 60 Garden St., Cambridge, MA 02138, USA}

\author[0000-0003-4784-3040]{Viviana V. Guzm\'{a}n}
\affiliation{Instituto de Astrof\'isica, Pontificia Universidad Cat\'olica de Chile, Av. Vicu\~na Mackenna 4860, 7820436 Macul, Santiago, Chile}

\author[0000-0003-1413-1776]{Charles J. Law}
\affiliation{Center for Astrophysics \textbar\ Harvard \& Smithsonian, 60 Garden St., Cambridge, MA 02138, USA}

\author[0000-0002-8932-1219]{Ryan A. Loomis}\affiliation{National Radio Astronomy Observatory, 520 Edgemont Rd., Charlottesville, VA 22903, USA}

\author[0000-0002-2700-9676]{Gianni Cataldi}
\affil{National Astronomical Observatory of Japan, Osawa 2-21-1, Mitaka, Tokyo 181-8588, Japan}
\affil{Department of Astronomy, Graduate School of Science, The University of Tokyo, Tokyo 113-0033, Japan}

\author[0000-0003-4001-3589]{Arthur D. Bosman}
\affiliation{Department of Astronomy, University of Michigan, 323 West Hall, 1085 S. University Avenue, Ann Arbor, MI 48109, USA}

\author[0000-0003-3283-6884]{Yuri Aikawa}
\affiliation{Department of Astronomy, Graduate School of Science, The University of Tokyo, Tokyo 113-0033, Japan}

\author[0000-0003-2253-2270]{Sean M. Andrews} 
\affiliation{Center for Astrophysics \textbar\ Harvard \& Smithsonian, 60 Garden St., Cambridge, MA 02138, USA}

\author[0000-0003-4179-6394]{Edwin A. Bergin} \affiliation{Department of Astronomy, University of Michigan, 323 West Hall, 1085 S. University Avenue, Ann Arbor, MI 48109, USA}

\author[0000-0003-2014-2121]{Alice S. Booth} \affiliation{Leiden Observatory, Leiden University, 2300 RA Leiden, the Netherlands}
\affiliation{School of Physics and Astronomy, University of Leeds, Leeds, UK, LS2 9JT}

\author[0000-0003-2076-8001]{L. Ilsedore Cleeves} \affiliation{Department of Astronomy, University of Virginia, 530 McCormick Rd, Charlottesville, VA 22904}

\author[0000-0002-1483-8811]{Ian Czekala}
\altaffiliation{NASA Hubble Fellowship Program Sagan Fellow}
\affiliation{Department of Astronomy \& Astrophysics, 525 Davey Laboratory, The Pennsylvania State University, University Park, PA 16802, USA}
\affiliation{Center for Exoplanets \& Habitable Worlds, 525 Davey Laboratory, The Pennsylvania State University, University Park, PA 16802, USA}
\affiliation{Center for Astrostatistics, 525 Davey Laboratory, The Pennsylvania State University, University Park, PA 16802, USA}
\affiliation{Institute for Computational \& Data Sciences, The Pennsylvania State University, University Park, PA 16802, USA}
\affiliation{Department of Astronomy, 501 Campbell Hall, University of California, Berkeley, CA 94720-3411, USA}

\author[0000-0001-6947-6072]{Jane Huang}
\altaffiliation{NASA Hubble Fellowship Program Sagan Fellow}
\affiliation{Department of Astronomy, University of Michigan, 323 West Hall, 1085 S. University Avenue, Ann Arbor, MI 48109, USA}
\affiliation{Center for Astrophysics \textbar\ Harvard \& Smithsonian, 60 Garden St., Cambridge, MA 02138, USA}

\author[0000-0003-1008-1142]{John~D.~Ilee} \affil{School of Physics \& Astronomy, University of Leeds, Leeds LS2 9JT, UK}

\author[0000-0003-1837-3772]{Romane Le Gal}
\affiliation{Center for Astrophysics \textbar\ Harvard \& Smithsonian, 60 Garden St., Cambridge, MA 02138, USA}
\affiliation{IRAP, Universit\'{e} de Toulouse, CNRS, CNES, UT3, 31400 Toulouse, France}
\affiliation{IPAG, Universit\'{e} Grenoble Alpes, CNRS, IPAG, 38000 Grenoble, France}
\affiliation{IRAM, 300 rue de la piscine, F-38406 Saint-Martin d'H\`{e}res, France}

\author[0000-0002-7607-719X]{Feng Long}
\affiliation{Center for Astrophysics \textbar\ Harvard \& Smithsonian, 60 Garden St., Cambridge, MA 02138, USA}

\author[0000-0002-7058-7682]{Hideko Nomura}\affiliation{National Astronomical Observatory of Japan, Osawa 2-21-1, Mitaka, Tokyo 181-8588, Japan}

\author[0000-0002-1637-7393]{Fran\c cois M\'enard}\affiliation{Univ. Grenoble Alpes, CNRS, IPAG, F-38000 Grenoble, France}

\author[0000-0001-8642-1786]{Chunhua Qi} \affiliation{Center for Astrophysics \textbar\ Harvard \& Smithsonian, 60 Garden St., Cambridge, MA 02138, USA}

\author[0000-0002-6429-9457]{Kamber R. Schwarz}
\altaffiliation{NASA Hubble Fellowship Program Sagan Fellow}
\affiliation{Lunar and Planetary Laboratory, University of Arizona, 1629 E. University Blvd, Tucson, AZ 85721, USA}

\author[0000-0003-1534-5186]{Richard Teague}
\affiliation{Center for Astrophysics \textbar\ Harvard \& Smithsonian, 60 Garden St., Cambridge, MA 02138, USA}

\author[0000-0002-6034-2892]{Takashi Tsukagoshi} \affiliation{National Astronomical Observatory of Japan, Osawa 2-21-1, Mitaka, Tokyo 181-8588, Japan}

\author[0000-0001-6078-786X]{Catherine Walsh}\affiliation{School of Physics and Astronomy, University of Leeds, Leeds, UK, LS2 9JT}

\author[0000-0003-1526-7587]{David J. Wilner}\affiliation{Center for Astrophysics \textbar\ Harvard \& Smithsonian, 60 Garden St., Cambridge, MA 02138, USA}

\author[0000-0003-4099-6941]{Yoshihide Yamato} \affiliation{Department of Astronomy, Graduate School of Science, The University of Tokyo, Tokyo 113-0033, Japan}

\begin{abstract}
\noindent UV photochemistry in the surface layers of protoplanetary disks dramatically alters their composition relative to previous stages of star formation. The abundance ratio CN/HCN has long been proposed to trace the UV field in various astrophysical objects, however to date the relationship between CN, HCN, and the UV field in disks remains ambiguous. As part of the ALMA Large Program MAPS (Molecules with ALMA at Planet-forming Scales), we present observations of CN N=1--0 transitions at 0.3'' resolution towards five disk systems.  All disks show bright CN emission within $\sim$50-150 au, along with a diffuse emission shelf extending up to 600 au.  In all sources we find that the CN/HCN column density ratio increases with disk radius from about unity to 100, likely tracing increased UV penetration that enhances selective HCN photodissociation in the outer disk.  Additionally, multiple millimeter dust gaps and rings coincide with peaks and troughs, respectively, in the CN/HCN ratio, implying that some millimeter substructures are accompanied by changes to the UV penetration in more elevated disk layers.  That the CN/HCN ratio is generally high ($>$1) points to a robust photochemistry shaping disk chemical compositions, and also means that CN is the dominant carrier of the prebiotically interesting nitrile group at most disk radii.  We also find that the local column densities of CN and HCN are positively correlated despite emitting from vertically stratified disk regions, indicating that different disk layers are chemically linked.  This paper is part of the MAPS special issue of the Astrophysical Journal Supplement.

\end{abstract}


\section{Introduction}
\label{sec:intro}
The compositions of planets and planetesimals are determined by the chemical inventory of the protoplanetary disk in which they form.  In turn, the disk composition is likely set by a combination of inheritance from the interstellar medium and in situ chemistry.  Photochemistry plays an important role in driving disk chemistry via the formation of ions and radicals in the strongly irradiated disk surface layer and the somewhat shielded molecular layer \citep[e.g.][]{Aikawa1999}. Thus, the radiation field of the disk is expected to strongly influence the degree to which in situ chemistry contributes to the overall disk composition.

UV radiation is predicted to play an especially important role, particularly in the somewhat elevated disk layers typically probed by ALMA \citep[e.g.][]{Cleeves2013}.  Tracing the UV field in disks is therefore a compelling goal in order to predict the extent to which in situ photochemistry is active within the disk.  The abundance ratio of CN/HCN has long been proposed as a tracer of the UV field, with high CN/HCN ratios originally identified towards UV-exposed planetary nebulae and photodissociation regions \citep[e.g.][]{Cox1992, Fuente1993, Fuente1995}.  The theoretical motivation for this is rooted in the different UV photodissociation cross-sections of both molecules.  While CN can only be efficiently photodissociated at UV wavelengths below $\sim$113~nm, HCN photodissociation cross-sections remain high up to $\sim$150~nm \citep[][]{Heays2017}.  Correspondingly, HCN is photodissociated at a faster rate than CN.  

In the context of protoplanetary disks, early chemical models predicted that, as a result of continuum UV exposure, CN would occupy lower-density, more UV-exposed radial and vertical regions of the disk compared to HCN \citep[e.g.][]{Aikawa1999, vanZadelhoff2003}. \citet{Bergin2003} were the first to highlight the importance of Ly-$\alpha$ radiation in setting the CN/HCN ratio in disks: Ly-$\alpha$ will selectively dissociate HCN, and in many cases the Ly-$\alpha$ line contains tens of percent of the total UV flux \citep[e.g.][]{Schindhelm2012}.  Subsequent modeling efforts have indeed supported that Ly-$\alpha$ plays a major role in setting the CN/HCN balance \citep{Fogel2011, Chapillon2012}.  These newer models similarly predicted that the CN/HCN ratio and the radial and vertical distributions of both molecules are sensitive to differential photodissociation.  However, they emphasized that high CN/HCN ratios in strong UV fields primarily reflect the loss of HCN, since CN formation via HCN photodissociation is typically minor compared to other CN formation pathways (i.e.\ reactions of N with CH or C$_2$).  

Observationally, high CN/HCN ratios ($\gtrsim$1) were identified in the first molecular surveys of disks and interpreted as a signature of active photochemistry \citep[e.g.][]{Dutrey1997, Kastner1997}.  Subsequent observational programs confirmed high CN/HCN ratios towards numerous protoplanetary disks relative to earlier stages of star formation \citep{Thi2004, Kastner2008, Salter2011, Oberg2010, Oberg2011, Chapillon2012}.  However, contrary to expectations, surveys including disks with widely varying physical properties revealed remarkably similar CN/HCN ratios \citep{Kastner2008, Salter2011, Oberg2011}, at odds with a scenario in which the CN/HCN ratio is set primarily by the strength of the UV field in a given disk.  

Still, these conclusions are based on line flux ratios rather than column density or abundance ratios, introducing possible degeneracies with line optical depth and excitation temperature.  HCN at least has been found to exhibit optically thick emission in many disks \citep[e.g.][]{Bergner2019}, which could obscure trends in the CN/HCN flux ratios.  Moreover, these previous observations were either single dish (i.e.~spatially unresolved) or only marginally resolved, hindering an exploration on the radial variation of CN/HCN ratios.  Indeed, \citet{Oberg2011} noted a differentiation in the radial extent of CN compared to HCN in two disks observed with the SMA, and proposed that radial trends in the CN/HCN ratio may be a more powerful tracer of the UV field compared to the disk-averaged value.  Similarly, \citet{Guzman2015} found a larger radial extent of CN than HCN in the disk around the Herbig Ae star MWC 480, but not the disk around the T Tauri star DM Tau.  Overall, the utility of the CN/HCN ratio as a UV tracer (both within a given disk and across disks) remains unclear.

The relationship between UV exposure and CN/HCN is further complicated by other UV-driven chemistry that has been proposed to impact CN and HCN formation and destruction. Recent modeling by \citet{Visser2018} and \citet{Cazzoletti2018} demonstrated the importance of CN formation in disks via FUV pumping of H$_2$ into a vibrationally excited state, termed H$_2^*$ \citep{Stecher1972, Tielens1985}.  The reaction between H$_2^*$ with N to form NH, followed by reaction with C or C$^+$, was found to be the dominant formation pathway for CN.  In this model, H$_2$ pumping was assumed to be proportional to its photodissociation rate, meaning that pumping of H$_2$ will be driven primarily by UV photons with wavelengths $<$105~nm \citep{Heays2017}.  \citet{Cazzoletti2018} showed that this framework successfully reproduces the ringed morphology of CN towards TW Hya \citep{Teague2016}, and later the same model roughly reproduced the CN emission towards two disks in the Lupus star-forming region \citep{vanTerwisga2019}.  An excitation analysis of CN emission in TW Hya \citep{Teague2020} found that CN emits from moderately elevated disk regions consistent with a UV-driven formation pathway.  Observations of CN emission towards the edge-on Flying Saucer disk also showed CN emission from elevated layers in the outer disk \citep{Ruiz2021}.  However, in some tension with the scheme outlined in \citet{Cazzoletti2018}, \citet{Arulanantham2020} found a negative correlation between FUV continuum fluxes and sub-millimeter CN fluxes in a sample of disks, which was interpreted as efficient CN photodissociation in sources with higher FUV fields.  The same authors also saw a positive correlation between the FUV continuum and IR HCN emission, implying UV-driven HCN formation chemistry in the inner disk.

In summary, the relationship between the UV field and HCN and CN formation and destruction may involve (i) selective photo-dissociation of HCN over CN by UV continuum ($>$113~nm) and Ly-$\alpha$ radiation, (ii) enhanced formation of CN in strong FUV fields ($<$105~nm) due to FUV pumping of H$_2^*$ as an initiator of CN formation, and (iii) enhanced destruction of CN and formation of HCN in high FUV fields.  

With this in mind, our objective is to explore the extent to which CN and HCN emission morphologies, column densities, and column density ratios trace the UV field in protoplanetary disks.  The high sensitivity and high spatial resolution of the MAPS data set \citep{Oberg2020} enables us to explore in detail the radial trends in the CN/HCN ratio in a sample of five physically diverse protoplanetary disks.  In Section \ref{sec:obs}, we introduce the ALMA observations and present moment zero maps and radial intensity profiles.  In Section \ref{sec:columns}, we derive radial column density profiles for CN and HCN, and search for correlations across the sample.  In Section \ref{sec:toymodel}, we use a toy model to explore connections between our observations and the MAPS disk UV fields.  In Section \ref{sec:discussion}, we discuss the implications for CN and HCN formation and destruction chemistry, and the utility of these molecules as photochemical tracers.  We summarize our conclusions in Section \ref{sec:conclusion}.

\begin{deluxetable*}{lcccccccccc}
	\tabletypesize{\footnotesize}
	\tablecaption{CN and HCN spectral line data \label{tab:linedat}}
	\tablecolumns{10} 
	\tablewidth{\textwidth} 
	\tablehead{
		\multicolumn{6}{c}{Transition}         &
		\colhead{Frequency}       & 
		\colhead{$E_u$}              &
		\colhead{log$_{10}$($A_{ul}$)}        &
		\colhead{$g_u$}              \\
		\colhead{N'}         &
		\colhead{N}         &
		\colhead{J'}         &
		\colhead{J}         &
		\colhead{F'}         &
		\colhead{F}         &
		\colhead{(GHz)}               & 
		\colhead{(K)}                        & 
		\colhead{(s$^{-1}$)}            &
		}
\startdata
\multicolumn{10}{c}{CN} \\ 
\hline 
1 & 0 & 3/2 & 1/2 & 3/2 & 1/2 & 113.488120 & 5.4 & $-$5.17 & 4 \\ 
1 & 0 & 3/2 & 1/2 & 5/2 & 3/2 & 113.490970 & 5.4 & $-$4.92 & 6 \\ 
1 & 0 & 3/2 & 1/2 & 1/2 & 1/2 & 113.499644 & 5.4 & $-$4.97 & 2 \\ 
1 & 0 & 3/2 & 1/2 & 3/2 & 3/2 & 113.508907 & 5.4 & $-$5.28 & 4 \\ 
\hline 
2 & 1 & 3/2 & 1/2 & 3/2 & 3/2 & 226.632190 & 16.3 & $-$4.37 & 4 \\ 
2 & 1 & 3/2 & 1/2 & 5/2 & 3/2 & 226.659558 & 16.3 & $-$4.02 & 6 \\ 
2 & 1 & 3/2 & 1/2 & 1/2 & 1/2 & 226.663693 & 16.3 & $-$4.07 & 2 \\ 
2 & 1 & 3/2 & 1/2 & 3/2 & 1/2 & 226.679311 & 16.3 & $-$4.28 & 4 \\ 
\hline 
\multicolumn{10}{c}{HCN} \\ 
\hline 
 & & 3 & 2 & 3 & 3 & 265.884891 & 25.5 & $-$4.03 & 7 \\ 
 & & 3 & 2 & 2 & 1 & 265.886189 & 25.5 & $-$3.15 & 5 \\ 
 & & 3 & 2 & 3 & 2 & 265.886434 & 25.5 & $-$3.13 & 7 \\ 
 & & 3 & 2 & 4 & 3 & 265.886500 & 25.5 & $-$3.08 & 9 \\ 
 & & 3 & 2 & 2 & 3 & 265.886979 & 25.5 & $-$5.43 & 5 \\ 
 & & 3 & 2 & 2 & 2 & 265.888522 & 25.5 & $-$3.89 & 5 \\ 
\enddata
\tablenotetext{}{Line parameters are taken from the CDMS database, with CN measurements provided by \citet{Dixon1977} and \citet{Skatrud1983} and HCN measurements from \citet{Ahrens2002}.}
\end{deluxetable*}

\section{Observations}
\label{sec:obs}
\subsection{Observational details}

The MAPS sample is composed of the T Tauri disk systems AS 209, GM Aur, and IM Lup, and the Herbig Ae disk systems HD 163296 and MWC 480.  GM Aur exhibits a central dust and gas cavity, while all other disks are full disks, though with a wide range of dust substructures \citep{Andrews2018, Long2018, Huang2018, Huang2020}.  Further source details can be found in MAPS I \citep{Oberg2020}.  Four hyperfine components of the CN N=1--0 transition were covered within the B3-2 setup of MAPS, which is described in detail in MAPS I.  All five disks were observed between October 2018 and September 2019. For all disks, combined short- and long-baseline executions resulted in a total range of baselines from 15--3638~m.  The spectral window containing the CN transitions has a channel spacing of 61 kHz, corresponding to $\sim$0.16 km s$^{-1}$ at 113 GHz.  

We also make use of the HCN J=3--2 hyperfine transitions covered within the B6-2 setup of MAPS.  We note that the MAPS HCN results are presented in full in MAPS VI \citep{Guzman2020}.  

Additionally, we make use of observations from the ALMA project 2016.1.00884.S (PI: V. Guzm\'an) which cover four hyperfine components of the CN N=2--1 transition towards HD 163296.  These observations were taken in three execution blocks between Nov 2016 and March 2017, with a total time on source of 67 minutes.  Baselines ranged from 15--1040~m.  The CN transitions are contained in a spectral window with a channel spacing of 122 kHz, corresponding to $\sim$0.16 km s$^{-1}$ at 226 GHz.

Spectral line information for the CN and HCN transitions considered in this work can be found in Table \ref{tab:linedat}.  Imaging of all lines was performed according to the MAPS imaging pipeline, as described in MAPS II \citep{Czekala2020}.  The resulting CN and HCN images have circularized beams with a FWHM of 0.3'' and 0.15'', respectively.  We note that for this project we produced CN N=1--0 images with a spectral resolution of 0.2 km s$^{-1}$, i.e. close to the native resolution, as opposed to the default 0.5 km s$^{-1}$ gridding used for Band 3 lines in MAPS I.  Thus, the Jorsater \& van Moorsel (JvM) correction factors $\epsilon$ \footnote{This corrects for non-Gaussian behavior of the dirty beam that results from combining short- and long-baseline observations; see  \citet{Jorsater1995} and \citet{Czekala2020} for details.} presented in Table 11 of MAPS I are still applicable, but the channel rms values for our images (across 0.2 km/s) are instead 0.94, 1.6, 0.87, 1.1, and 1.6 mJy/beam towards AS 209, GM Aur, HD 163296, IM Lup, and MWC 480, respectively.  

For the CN N=2--1 observations towards HD 163296, the restoring beam dimensions are 0.5$\times$0.5''.  These observations combine data from two different ALMA configurations, and so we apply the JvM correction with an $\epsilon$ of 0.704. The channel rms in 0.2 km/s channels is 2.9 mJy/beam.

\subsection{Observational results}
\label{subsec:obs_morphologies}

Here we present the MAPS CN observations, including moment zero maps and radial intensity profiles.  For HCN, similar figures are presented in MAPS VI \citep{Guzman2020}.  Moment zero maps were constructed as described in MAPS III \citep{Law2020_rad}.  We emphasize that these images are presentational only, and all quantitative analysis is performed on the full image cubes.  In the case of CN N=1--0, the J=$\frac{3}{2}$--$\frac{1}{2}$, F=$\frac{5}{2}$--$\frac{3}{2}$ hyperfine component is the brightest, and therefore the target of our morphological analysis.  However, this line is blended with the J=$\frac{3}{2}$--$\frac{1}{2}$, F=$\frac{3}{2}$--$\frac{1}{2}$ component.  We therefore construct moment zero maps from the combined emission of both hyperfine components, to prevent artificial emission asymmetries. 

\begin{figure*}
    \includegraphics[width=\linewidth]{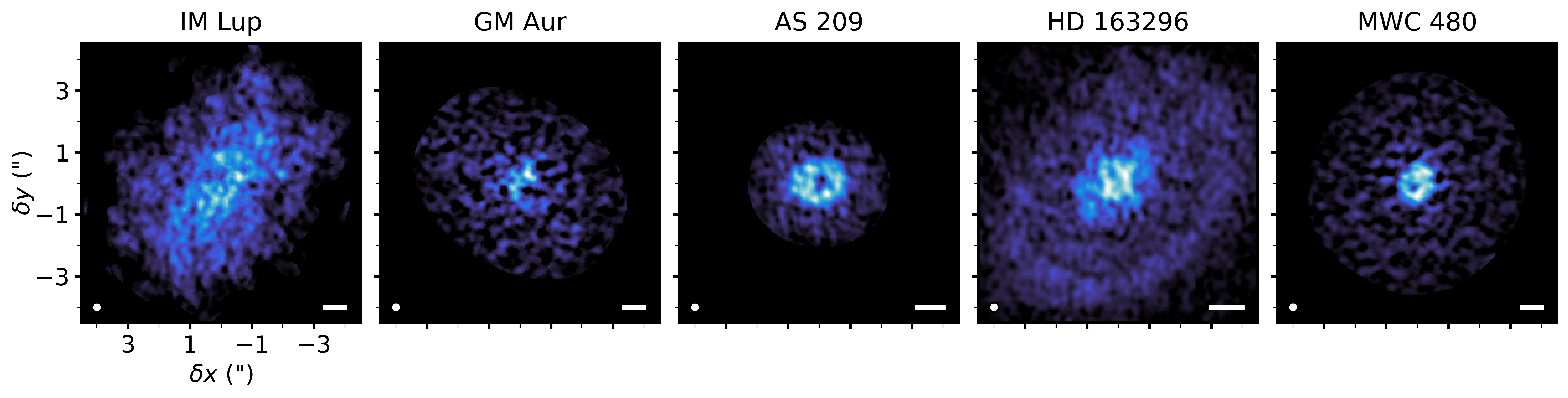}
    \caption{Presentational moment zero maps of CN N=1--0 emission (including the two hyperfine components at 113.488 and 113.491 GHz) towards the five MAPS disks. To better illustrate weak emission features, integrated intensities are shown with a power-law color stretch.  Color scales are normalized to each individual image.  In each panel, the restoring beam is shown in the lower left and a 100 au scale bar is shown in the lower right.}
    \label{fig:CN_mom0}
\end{figure*}

The resulting moment zero maps for the combined CN N=1--0, J=$\frac{3}{2}$--$\frac{1}{2}$, F=$\frac{5}{2}$--$\frac{3}{2}$ and $\frac{3}{2}$--$\frac{1}{2}$ lines are shown in Figure \ref{fig:CN_mom0} for the five disks in the MAPS survey.  Figure \ref{fig:mom0_cn21} in Appendix \ref{sec:app_columns} shows the moment zero map for the CN N=2--1, J=$\frac{3}{2}$--$\frac{1}{2}$, F=$\frac{5}{2}$--$\frac{3}{2}$ and $\frac{1}{2}$--$\frac{1}{2}$ transitions towards HD 163296.

\begin{figure*}[t]
    \includegraphics[width=\linewidth]{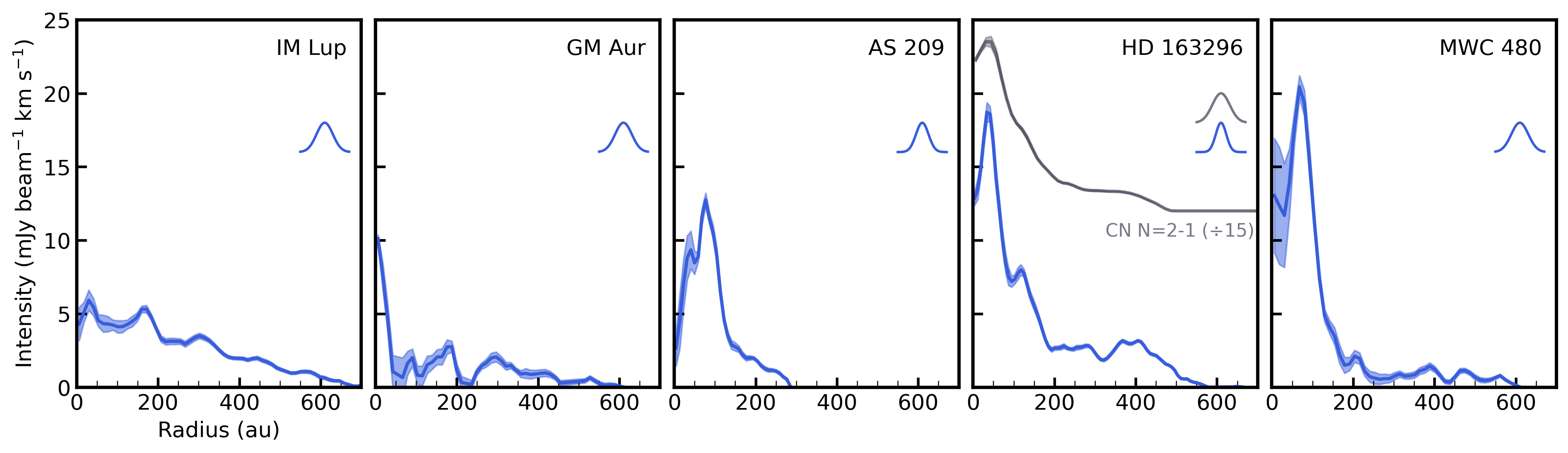}
    \caption{CN N=1--0 deprojected radial intensity profiles, including only pixels within a $\pm$30$^\circ$ wedge along the disk major axis.  CN N=2--1 is also shown for HD 163296  (grey line), scaled down by a factor of 15 and offset for clarity.  Shaded regions represent the uncertainties in each radial bin, as described in \citet{Law2020_rad}.  The synthesized beams are shown in the upper right of each panel.}
    \label{fig:CN_radprofs}
\end{figure*}

Deprojected radial intensity profiles were generated as described in MAPS III \citep{Law2020_rad}, and are shown in Figure \ref{fig:CN_radprofs}.  We included only pixels within a $\pm$30$^\circ$ wedge along the disk major axis in order to avoid the disk regions most impacted by potential flaring, which can obscure the presence of substructures. 

As seen in the moment zero maps and radial intensity profiles, the IM Lup disk shows a small central emission dip followed by a small emission peak within 50 au, additional small peaks around 180 and 320 au, and an extended emission shelf out to 600 au.  GM Aur, the only disk without a central emission dip, shows centrally concentrated CN emission within 50 au, though we cannot rule out that an inner emission hole would be detected at higher spatial resolution.  The GM Aur disk exhibits weak emission out to $\sim$600 au, with notable bumps around 200 and 300 au, and a gap around 220 au.  The AS 209 disk shows a deep CN emission hole, surrounded by a bright and fairly compact ring of emission within $\sim$150 au that appears to consist of two components, and may resolve into two separate rings at higher resolution.  Beyond this, there is a weaker emission shelf out to $\sim$300 au.  The HD 163296 disk shows a central dip in CN emission followed by a bright ring around 50 au, with another peak around 150 au.  Beyond this there is relatively flat emission from $\sim$200-300 au, followed by a ring centered around 400 au and extending to 600 au.  These features can be seen both in the CN N=1--0 (lower SNR, higher resolution) and 2--1 (higher SNR, lower resolution) images.  Towards MWC 480, the CN emission appears to exhibit a moderate central emission dip, though the uncertainties are high.  From the moment zero map it appears that the emission hole is present only on one side of the disk.  Higher resolution observations are needed to better determine the CN morphology in the inner $\sim$50 au of the MWC 480 disk.  Beyond this, there is a bright and compact ring out to 150 au, and a diffuse emission component extending to $\sim$600 au. Small-scale bumps are seen at 200 au, 400 au, and 450 au, with a gap at 420 au, though these features are weak.  Indeed, several of the weak emission bumps in IM Lup, GM Aur, and MWC 480 are not distinct from the surrounding line profile at a $>$3-$\sigma$ level, and we advise caution in interpreting these features until higher-SNR observations can confirm that they are real.  When considering only the unambiguous emission features, we conclude that most disks show a bright and compact emission ring within $\sim$50--150 au (though the GM Aur disk shows a bright emission peak with no inner depression), and weaker emission rings/shelves extending out to a few hundred au. 

\section{CN and HCN column densities}
\label{sec:columns}
\subsection{Spectral line model}
\label{subsec:hf_fitting}
Both CN and HCN exhibit hyperfine splitting of their rotational spectra, which can be fitted in order to obtain constraints on their column densities.  Our modeling framework is as follows.  We produce synthetic spectra assuming that all hyperfine components share a common rotational temperature $T_{r}$ and that the emission is in local thermodynamic equilibrium (LTE).  The critical densities are below 10$^5$ cm$^{-3}$ for the observed CN transitions and below a few$\times10^7$ cm$^{-3}$ for the observed HCN transitions.  Based on the disk structure models derived in \citet{Zhang2020}, disk densities exceed the CN critical density for typical CN emission heights \citep[$z/r \sim$0.2;][]{Teague2020}. Non-LTE modeling of HCN in the MAPS disks by \citet{Cataldi2020} shows that while the HCN emitting region may drop below the critical density in the outer disk, the impact on column density is small in the disk regions that we are focused on (i.e. within $\sim$400 au).  

For each hyperfine component $i$, the optical depth at the line center $\tau_{i,0}$ is found from:
\begin{equation}
    \tau_{i,0} = \frac{N_T}{Q(T_{r})} e^{-E_{u,i}/T_{r}} \frac{g_{u,i} A_{u, i}c^3}{8\pi\nu_i^3}\frac{1}{\sigma \sqrt{2\pi}} (e^{h\nu_i/kT_{r}}-1),
    \label{eq:tau}
\end{equation}
where $N_T$ is the total column density, $Q(T_{r})$ is the molecular partition function, $\sigma$ is the Gaussian line width, and for each hyperfine component, $E_{u,i}$ is the upper state energy (K), $g_{u,i}$ is the upper state degeneracy, $A_{u,i}$ is the Einstein coefficient and $\nu_i$ is the transition frequency.

We assume Gaussian line profiles such that the total optical depth profile is:
\begin{equation}
    \tau_\nu = \sum_{i}\tau_{i,0} \mathrm{exp}\Big{(}\frac{-(V-V_i - V_{lsr})^2}{2\sigma^2}\Big{)},
    \label{eq:tau_nu}
\end{equation}
where $V$ is the velocity, $V_i$ is the velocity offset of each hyperfine component, and $V_{lsr}$ is the systemic velocity.

Lastly, we solve for the intensity profile (in units Jy/beam) with:
\begin{equation}
    I_\nu = [B_\nu(T_{r}) - B_\nu(T_{bg})] \times (1-e^{-\tau_v}) \times \Omega,
    \label{eq:F_nu}
\end{equation}
where $B_\nu$ is the Planck function and $\Omega$ is the effective angular area of the restoring beam.  For the background temperature $T_{bg}$, we adopt the maximum of the cosmic microwave background (2.73 K) or the average continuum brightness temperature within the same radial bin as the line data.  We use the continuum images from the MAPS setup B3-2 (2.8 mm) and B6-2 (1.2 mm) for CN and HCN, respectively, which are shown in Figure 1 of MAPS XIV \citep{Sierra2020}.  

In principle, the line width should be dominated by thermal broadening, as turbulence has been found to be minor in disks \citep[e.g][]{Teague2016, Flaherty2020}.  In practice, the line widths measured in our spectra are broader than thermal.  This observational broadening could be due to a combination of (1) emission contributions from the back side and the front side of a flared disk, which have different projected velocities, (2) beam smearing, in which a wide range of Keplerian velocities are incorporated within the same beam, especially impacting the inner disk, and (3) Hanning smoothing of the data which effectively reduces the spectral resolution.  Because the optical depth is dependent on the line width (Equation \ref{eq:tau}), this extra line broadening will result in artificially low inferred optical depths. 

To correct for this, we adopt the thermal line width $\sigma_{\mathrm{therm}}$ as the line width when calculating $\tau_{i,0}$ from Equation \ref{eq:tau}, as a better approximation of the excitation conditions within a given radial bin than what is inferred from the broadened line.  As noted above, we do not expect turbulence to be an important contributor to the line width \citep[e.g.][]{Teague2016,Flaherty2020}.   The thermal line width is found from:
\begin{equation}
    \sigma_{\mathrm{therm}} = \sqrt{\frac{kT_{kin}}{m}},
\end{equation}
where $m$ is the molecule's mass.  We adopt the rotational temperature as the kinetic temperature $T_{kin}$, i.e. assuming that the emission is well described by LTE.  With $\tau_{i,0}$ calculated assuming $\sigma = \sigma_{\mathrm{therm}}$ (Equation \ref{eq:tau}), we then calculate $\tau_\nu$ and $I_\nu$ according to Equations \ref{eq:tau_nu} and \ref{eq:F_nu}.  Lastly, the total integrated intensity of the `thermal' spectrum is redistributed assuming a Gaussian line profile with a broadened line width $\sigma_{\mathrm{obs}}$, preserving the total velocity-integrated intensity.

eWe fit for the free parameters $N_T$, $T_{r}$, $\sigma_{\mathrm{obs}}$, and V$_{lsr}$.  We use the MCMC package \texttt{emcee} \citep{Foreman-Mackey2013} to sample the posterior distributions of the four fit parameters.  

\subsection{Line fitting}

\begin{figure}
    \includegraphics[width=0.9\linewidth]{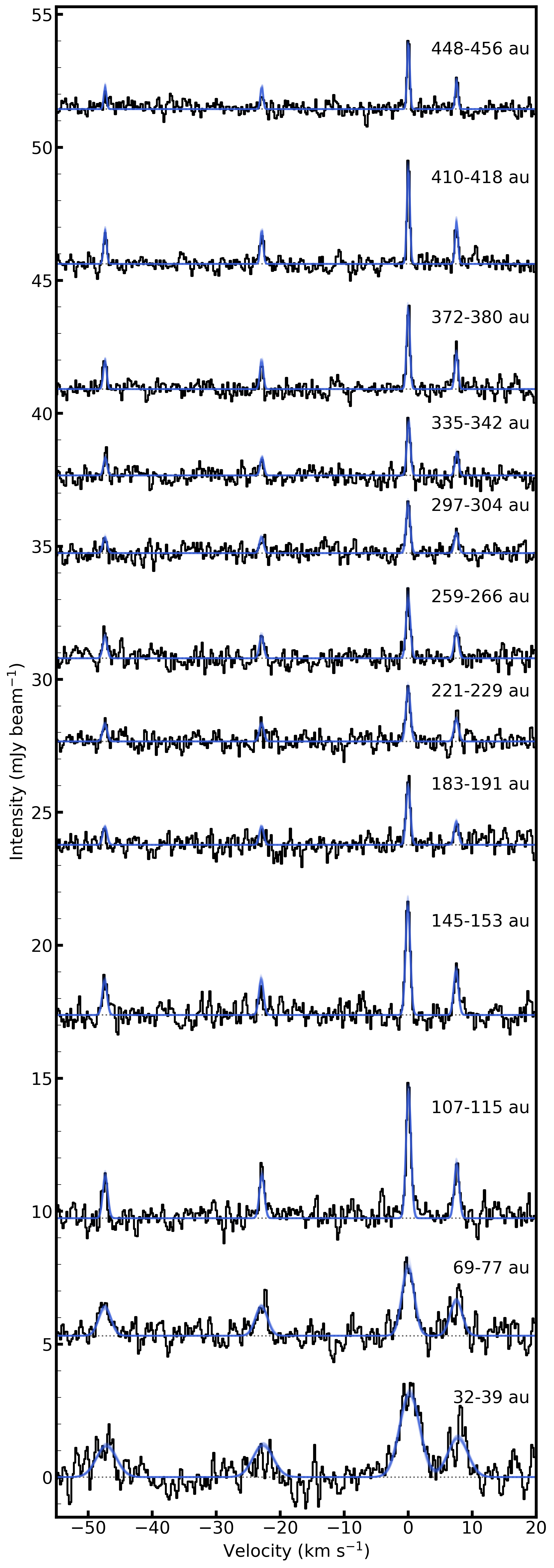}
    \caption{Example radial spectra of the CN N=1--0 hyperfine components in HD 163296 (black lines), averaged within a $\pm$30$^\circ$ wedge along the disk major axis.  Draws from the spectral line fit posteriors are shown in blue.  Spectra are offset for clarity, and the radial range of each spectrum is indicated on the right.}
    \label{fig:ex_spec_fits}
\end{figure}

\begin{figure*}
    \includegraphics[width=\linewidth]{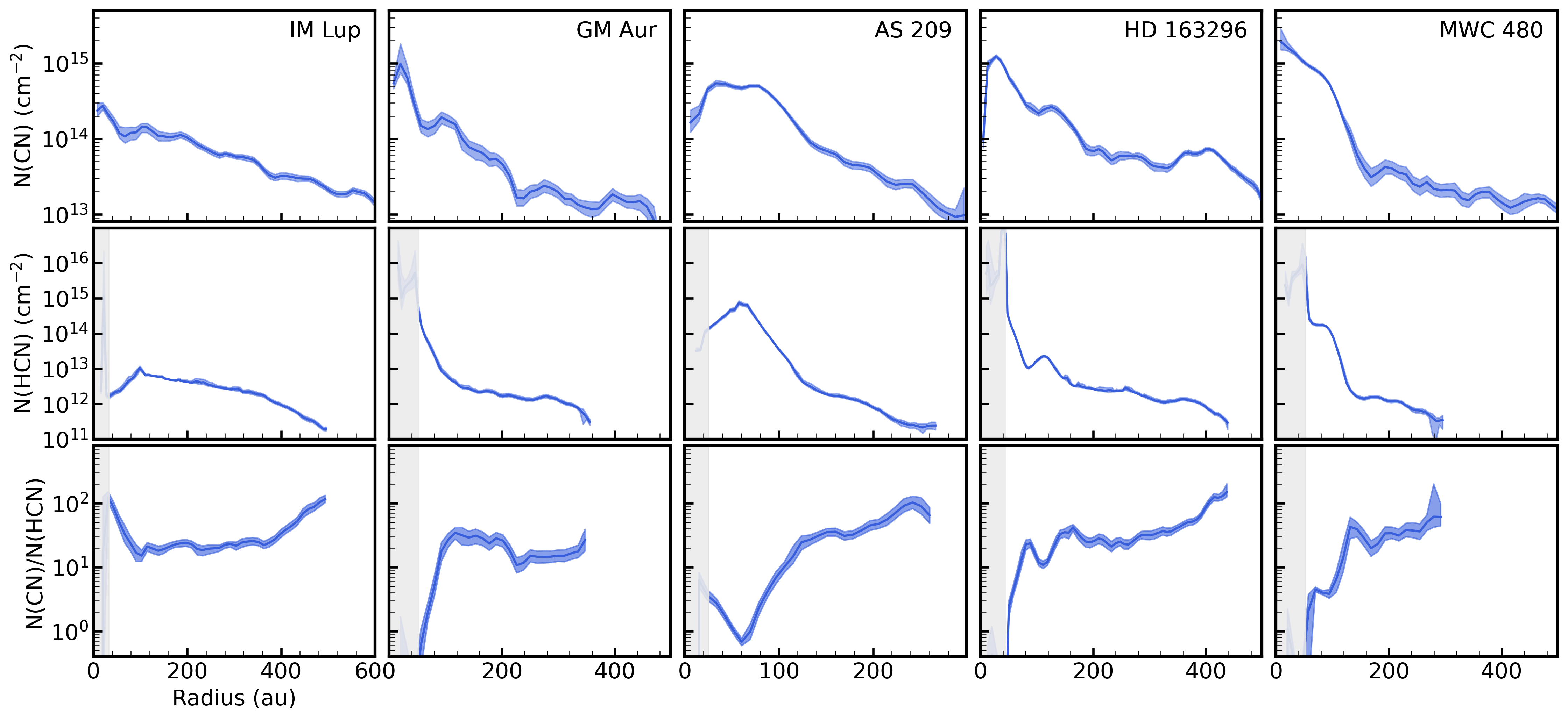}
    \caption{CN (top) and HCN (middle) column densities, and the CN/HCN column density ratios (bottom), derived from hyperfine fitting.  Solid blue lines represent the median and shaded blue regions represent the 16th-84th percentiles of the MCMC posteriors.  The greyed-out regions represent disk radii for which HCN line blending is too severe to allow reliable constraints on the column densities.}
    \label{fig:cn_hcn_columns}
\end{figure*}

We perform line fitting on radial spectra obtained using the task \texttt{cube.radial\_spectra} within the Python package \texttt{gofish} \citep{Teague2019}.  With this technique, pixels are corrected for their projected Keplerian velocity prior to averaging \citep[e.g.][]{Yen2016}, thus limiting extra line broadening which would normally be introduced in a radial average.  Spectra are generated in radial bins that are 1/4 of the beam size.  For lines with sufficient SNR (HCN towards all disks and CN towards HD 163296 and AS 209), we use only pixels within a $\pm$30$^\circ$ wedge along the disk major axis.  This allows us to more clearly identify column density substructures by minimizing distortions to the radial profile that are caused by flaring, an effect that is most impactful along the disk minor axis.  This treatment does not significantly change the derived column densities compared to azimuthal averages.  For CN in MWC 480, GM Aur, and IM Lup we include all pixels in the azimuthal averages in order to achieve sufficiently high SNRs to perform line fitting.  To extract the HCN radial spectra in HD 163296, we assume an emission surface of $z/r$=0.1 based on the fitting in MAPS IV \citep{Law2020_surf}.  For all other sources, we do not assume an emission surface since the surface either could not be constrained, or is consistent with a $z/r \lesssim$0.1.  We also note that the HCN column density profiles we derive are in very good agreement with those independently derived in MAPS VI \citep{Guzman2020} and MAPS X \citep{Cataldi2020}.

Figure \ref{fig:ex_spec_fits} shows example radial spectra of CN towards HD 163296, along with draws from the spectral model fit posteriors.  We note that because all four hyperfine components of the CN N=1--0 transition share a similar upper state energy, the derived excitation temperatures have large uncertainties.  To better inform our priors, we jointly fit the CN N=2--1 lines and 1--0 lines observed towards HD 163296, which resulted in tighter constraints on the CN excitation temperatures, and also confirmed that column density fits derived from the N=1--0 lines alone are indeed reliable.  Further details on the joint fitting can be found in Appendix \ref{sec:app_columns}.  

In the case of HCN, the hyperfine components are severely blended in the inner tens of au, prohibiting us from obtaining reliable column density constraints.  We therefore do not consider the HCN column densities for the affected radii, which are identified by-eye based on poorly converged walkers in the corner plots.  Detailed radiative transfer modeling of the inner disk HCN emission is needed to properly treat line blending and account for contributions from different surfaces in the disk.  We note that this is not a problem for the CN lines since the spacing between the hyperfine components is larger.   

The CN images used in our fitting have a 0.3'' resolution, while the HCN images have a 0.15'' resolution.  We chose to use the higher-resolution HCN images because there is less line blending between the HCN hyperfine components in higher spatial resolution data, thus we obtain more robust constraints on the column densities.  To obtain CN/HCN ratios, we draw 10,000 CN column density posteriors within a given CN radial bin and 10,000 HCN column density posteriors within the corresponding HCN radial bins, and find the 16th, 50th, and 84th percentile of CN/HCN ratios within this sample.

\subsection{CN and HCN column density profiles}
\label{subsec:col_profiles}

The CN and HCN column density profiles derived from hyperfine fitting are shown in Figure \ref{fig:cn_hcn_columns}, along with the CN/HCN column density ratios.  The rotational temperature and optical depth profiles are shown in Appendix \ref{sec:app_Tr_tau}.  Across the disk sample, the column densities range from 10$^{13}$--10$^{15}$cm$^{-2}$ for CN, and from 10$^{11}$--10$^{16}$cm$^{-2}$ for HCN (and possibly higher in the inner tens of au).  The range of CN column densities is quite narrow compared to HCN, perhaps implying formation in a limited range of physical environments.  

All disks except MWC 480 show an inner dip in the CN column density, though we note that there are large uncertainties on the CN intensity profile in the inner disk of MWC 480 due to the presence of an asymmetric emission dip (Figures \ref{fig:CN_mom0} and \ref{fig:CN_radprofs}).  Note also that unlike the radial intensity profile, the CN column density profile for GM Aur shows an inner dip.  However, the uncertainties are high due to broad line widths in the inner-most radial bin.  Higher spatial resolution observations are needed to determine the CN morphology in the inner disk of MWC 480 and GM Aur.  

In all disks, there is a ring of high CN column densities ($>$10$^{14}$~cm$^{-2}$) which extends to $\sim$50 au in IM Lup and GM Aur, and to $\sim$100--150 au in AS 209, HD 163296, and MWC 480.  Immediately beyond this, a second bump in column density is seen clearly for AS 209 and HD 163296, and may also be present in IM Lup, GM Aur, and MWC 480.  Further out, all disks exhibit CN shelves out to a few hundred au, with small-scale bumps and gaps mirroring those seen in the radial intensity profiles (Figure \ref{fig:CN_radprofs}).  However, we caution against over-interpreting these features due to the relatively low SNR of the CN emission in the outer disks.  In addition, HD 163296 exhibits a high-column density CN ring around 400 au.

The HCN column density profiles in IM Lup and AS 209 exhibit broad ringed profiles, consisting of a single ring component for IM Lup and an inner ring plus outer shelf in AS 209.  The column density profiles in the inner tens of au are poorly constrained for GM Aur, HD 163296, and MWC 480 due to line blending.  Around 50 au, where the fits become reliable, all three show a steeply decreasing column density profile.  HD 163296 shows a column density bump around 120 au, and MWC 480 exhibits a shoulder around 100 au.  All five disks show a gradual decline in the HCN column density beyond $\sim$150 au.  

\begin{figure*}
    \centering
    \includegraphics[width=0.9\linewidth]{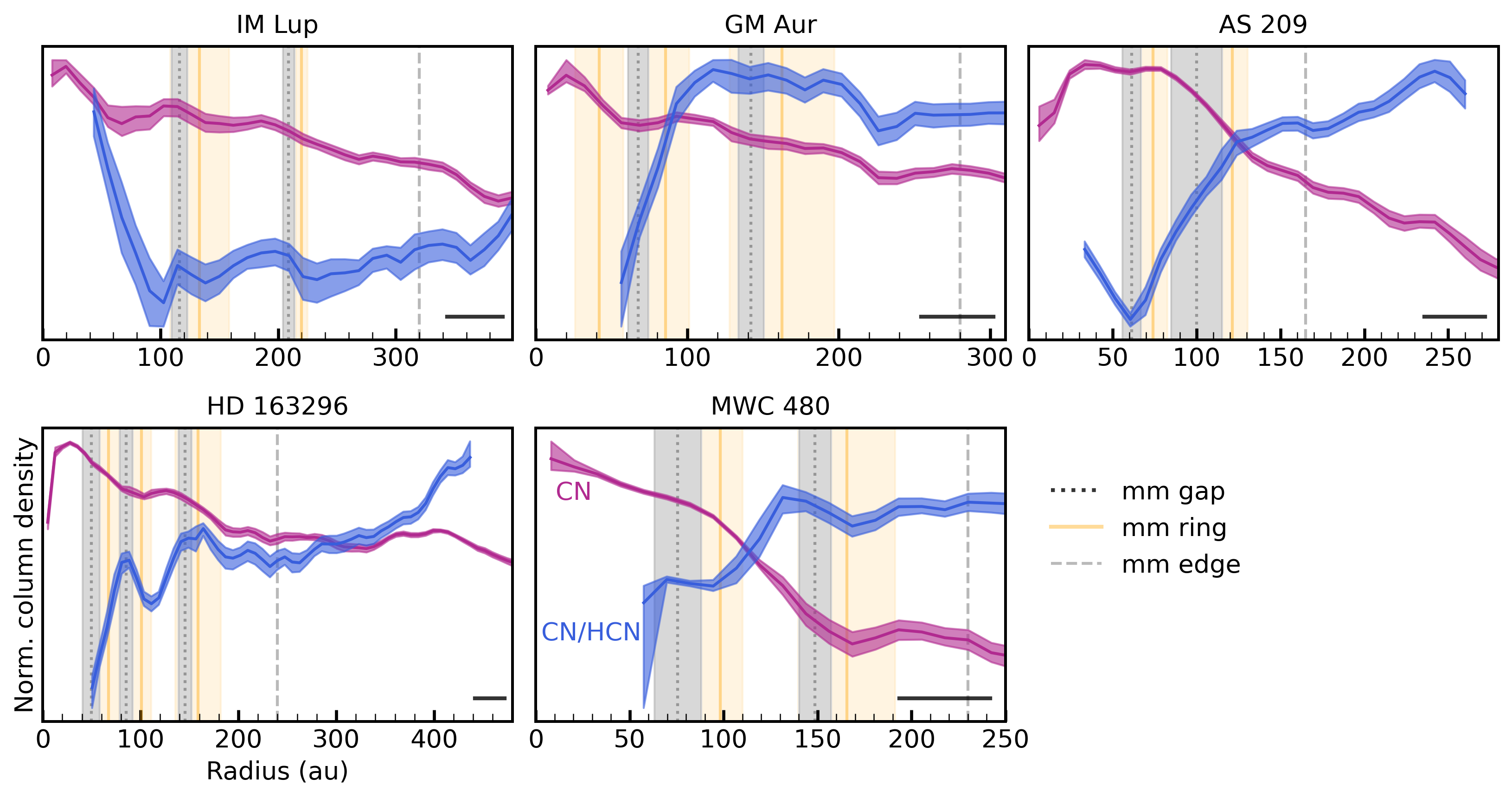}
    \caption{Normalized column density profiles of CN and CN/HCN (pink and blue lines, respectively) with millimeter dust features overplotted.  Solid orange and dotted grey vertical lines represent the positions of millimeter rings and gaps, respectively.  The substructure widths are indicated with shaded regions. The millimeter edge is shown as a dashed grey line.  Substructure properties are taken from \citet{Law2020_rad}.  Horizontal black lines show the restoring beam FWHM of the line data.}
    \label{fig:Nt_substructures}
\end{figure*}

The CN/HCN column density ratios are quite similar in GM Aur, AS 209, HD 163296, and MWC 480: a minimum CN/HCN occurs around 50 au, followed by a steep increase in CN/HCN out to 100--150 au, and a smoother increase in the outer disk.  In these disks, the CN/HCN ratio ranges from about unity at small radii to as high as 100 at larger radii.  IM Lup is distinct in that the minimum CN/HCN ratio is an order of magnitude higher than in the other disks, and the CN/HCN profile does not show the same dramatic increase around 50 au.  For the two disks with good constraints on the HCN column density in the inner disk, IM Lup and AS 209, we also observe that the CN/HCN ratio decreases steeply with radius in the inner 50 au.  Better constraints on the HCN column density at small scales is required to determine if this pattern is also present in GM Aur, HD 163296, and MWC 480.

In \citet{Guzman2015}, the CN emission was found to be radially extended relative to HCN in a disk around a Herbig star (MWC 480) but not a disk around a T Tauri star (DM Tau). In our sample, the radial extents of CN and HCN are comparable in AS 209, but the CN emission is more extended in the remaining disks.  It is interesting to note that both T Tauri systems (IM Lup and GM Aur) and Herbig Ae systems (HD 163296 and MWC 480) exhibit larger radial extents of CN compared to HCN.  Thus, our observations do not support any systematic difference between the relative CN vs.\ HCN extents around stars of different spectral types.

\subsection{Comparison to dust substructures}
\label{sec:line_vs_cont}

Figure \ref{fig:Nt_substructures} shows the normalized CN and CN/HCN column density profiles with millimeter dust substructure locations overplotted.  Millimeter substructure properties are taken from \citet{Law2020_rad}.  For clarity, we do not show HCN here since an analogous comparison of HCN column density with dust substructure is shown in \citet{Guzman2020} using the higher-resolution Band 6 data.

For the purposes of identifying possible associations between the line and continuum substructures, we consider cases in which the CN column density or CN/HCN ratio peaks or troughs either (i) within the width of the millimeter substructure, or (ii) within 0.075" from the center of the millimeter substructure, corresponding to a quarter of the CN beam FWHM.  We consider both criteria because the continuum substructure positions were derived using higher-resolution images than the CN images.  As a result, beam smearing could introduce an offset in the observed line peak position relative to the continuum.  This uncertainty would be mitigated by higher-resolution line observations to better identify the positions of CN and CN/HCN substructures.

There is no one-to-one trend between dust substructures and CN or CN/HCN substructures.  However, there are a number of cases in which the continuum intensity anti-correlates with the CN/HCN ratio.  This occurs for the two gap-ring pairs in IM Lup; for the ring-gap-ring sequence from 85-100 au in HD 163296; and for the inner gap-ring pair and outer ring in MWC 480.  A notable exception to this pattern is the large dip in CN/HCN at the position of the 61 au millimeter gap in AS 209.  For all other millimeter gaps and rings, there is either no corresponding feature in the CN/HCN ratio, or the relationship is ambiguous (as for the CN/HCN peaks around 160 au in HD 163296 and around 130 au in MWC 480).

An analogous overlay of the CN and CN/HCN column density profiles with NIR ring locations is shown in Figure \ref{fig:columns_nir}. Because NIR ring width measurements are not available, we consider an association if a line peak or trough is within a quarter beam width of a NIR ring.  We find that NIR rings are in some cases coincident with CN/HCN dips (92 and 152 au in IM Lup), in some cases with CN/HCN peaks (243 au in AS 209, 77 au in HD 163296), and in all remaining cases unassociated with any CN/HCN feature (281 and 332 au in IM Lup, 78 and 140 au in AS 209, 333 au in HD 163296).  Thus, it is difficult to identify a predictable change in CN/HCN ratio as a function of NIR features.

\subsection{Local CN vs.\ HCN column densities}

\begin{figure}
    \includegraphics[width=\linewidth]{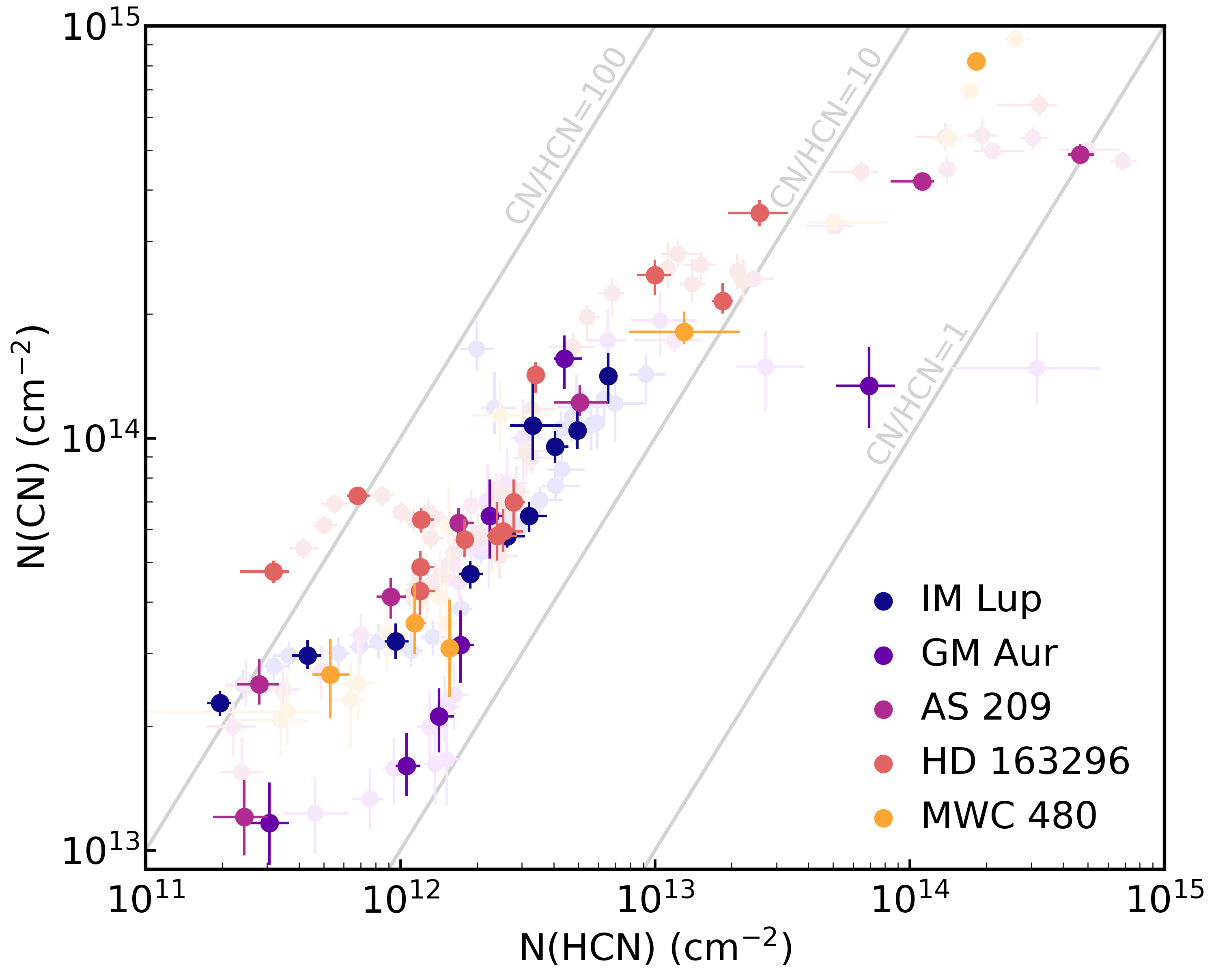}
    \caption{CN and HCN column densities for each radial bin in the five MAPS disks.  One radial bin per beam is shown as an opaque marker, and all other bins sampling the same beam are shown as transparent markers.  Constant CN/HCN ratios are marked with grey lines.}
    \label{fig:cn_vs_hcn}
\end{figure}

We use the radially resolved column density profiles to explore how the CN and HCN column densities are related to one another locally within a disk.  Figure \ref{fig:cn_vs_hcn} shows the CN and HCN column densities measured within each radial bin in the five MAPS disks.  We include only bins for which both CN and HCN column densities could be measured (see Figure \ref{fig:cn_hcn_columns}).  Bins that are redundant samples of the same beam are shown with transparent markers in order to limit artificial correlations.  

It is clear that local column densities of CN and HCN are positively correlated: disk radii with higher HCN column densities generally have higher CN column densities as well.  The molecular column densities appear to be well rank-ordered, but the proportionality constant changes such that the CN/HCN ratio decreases with increasing HCN column density.  Interestingly, points from different disks fall along roughly the same trend line, meaning that the CN column density can be inferred for a given HCN column density. Note that better constraints on the HCN column density in the inner disk regions are needed to determine if this trend is valid at all disk radii, or only beyond a few tens of au.  Still, the trend between CN and HCN column densities in a given disk location suggests that (i) the CN and HCN chemistries are linked, and (ii) their relative abundances are sensitive to the local disk environment.  We discuss this further in Section \ref{subsec:disc_corr}

\begin{figure*}
    \includegraphics[width=\linewidth]{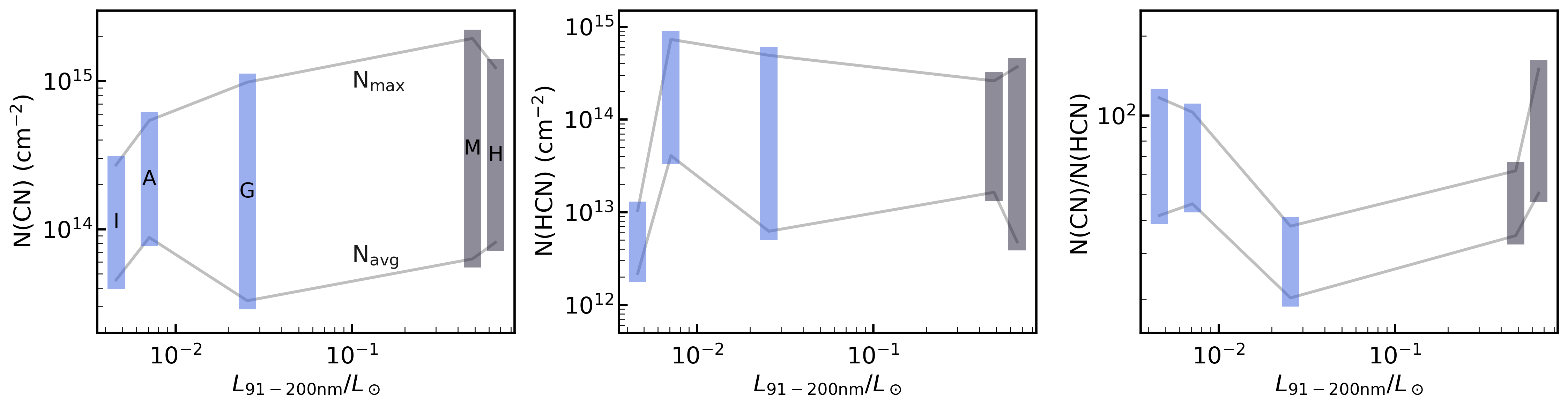}
    \caption{CN (left), HCN (middle), and CN/HCN (right) column densities in each disk plotted against the stellar UV luminosities.  The top of each bar represents the peak column density across all radii in a disk, and the bottom of each bar is the average column density in the disk. Each disk is labeled by its first letter.  Blue and grey bars represent T Tauri and Herbig Ae systems, respectively.}
    \label{fig:lum_columns}
\end{figure*}

\begin{deluxetable*}{lrrccc}
	\tabletypesize{\footnotesize}
	\tablecaption{Column density statistics \label{tab:columns}}
	\tablecolumns{6} 
	\tablewidth{\textwidth} 
	\tablehead{
		\colhead{Source}       & 
		\colhead{CN N$_{\mathrm{avg}}$ [N$_{\mathrm{max}}$]} &
		\colhead{HCN N$_{\mathrm{avg}}$ [N$_{\mathrm{max}}$]} &
	    \colhead{CN/HCN N$_{\mathrm{avg}}$ [N$_{\mathrm{max}}$]} &
		\colhead{R$_{\mathrm{CN}}$}         &
		\colhead{R$_{\mathrm{HCN}}$}         \\
		\colhead{}         &
		\colhead{(cm$^{-2}$)}         &
		\colhead{(cm$^{-2}$)}         &
		\colhead{(cm$^{-2}$)}         &
		\colhead{(au)}         &
		\colhead{(au)}         }
\startdata
IM Lup & 4.5 [27.1] $\times$10$^{13}$  & 2.2 [10.4] $\times$10$^{12}$  & 42 [117]  & 8--589 & 39--496\\ 
GM Aur & 3.3 [98.2] $\times$10$^{13}$  & 6.2 [491.6] $\times$10$^{12}$  & 20 [38]  & 8--493 & 52--356\\ 
AS 209 & 8.8 [54.1] $\times$10$^{13}$  & 4.1 [72.9] $\times$10$^{13}$  & 46 [103]  & 6--296 & 26--266\\ 
HD 163296 & 8.1 [123.2] $\times$10$^{13}$  & 4.8 [368.9] $\times$10$^{12}$  & 51 [150]  & 5--497 & 48--438\\ 
MWC 480 & 6.3 [194.1] $\times$10$^{13}$  & 1.6 [26.0] $\times$10$^{13}$  & 35 [62]  & 8--489 & 59--295\\ 
\enddata
\tablenotetext{}{Column density averages and maxima are found from the range of radii listed for each molecule.  CN/HCN ratios are derived using the HCN radii.}
\end{deluxetable*}

\subsection{Column density relationship to UV luminosity}
\label{subsec:col_uv}
As outlined in Section \ref{sec:intro}, CN and HCN formation and destruction are predicted to depend strongly on the local UV field.  Figure \ref{fig:lum_columns} shows the measured CN and HCN column densities, and CN/HCN ratios, plotted against the stellar UV luminosities (91-200 nm).  UV luminosities are calculated from the stellar spectra derived in MAPS V \citep{Zhang2020} based on the UV spectra presented in \citet{Dionatos2019}.  In order to explore trends between the stellar UV field and the bulk disk chemistry, we consider two column density metrics: the peak column density across all radial bins within a disk, and the average column density in all radial bins across a disk, weighted by the area of each bin.  These column density statistics are listed in Table \ref{tab:columns}.

As seen in Figure \ref{fig:lum_columns}, we observe no clear trends between the maximum or average column densities with the stellar UV luminosity.  In fact, the average CN column density varies by just a factor of $\sim$2.5 across the disk sample, and the peak by a factor of $\sim$7, despite two orders of magnitude difference in the stellar UV luminosities.  Thus, while a tentative increase in the maximum CN column density with UV luminosity can be seen, any dependency of bulk CN formation on stellar UV luminosity is quite weak.  HCN column densities exhibit a wider range of average and peak column densities, but show no clear trend with UV luminosity.  The average and peak CN/HCN ratios across the disks vary by factors of $\sim$2.5 and 4, respectively, and show no trend with stellar UV luminosity.  Thus, the CN/HCN ratio does not appear associated with the stellar UV luminosity.

It is important to note that the UV spectra used in this analysis include measurements from different instruments taken at different times, and the resulting co-added spectra may not be representative of the UV environment at the time of our ALMA observations.  Thus, it is possible that trends with stellar UV luminosities exist but are not identifiable with the current UV spectra.  Still, for CN and CN/HCN, any dependency is necessarily subtle due to the small ranges observed across the disk sample.

\subsection{CN rotational temperature profile in HD 163296}
\label{sec:CN_Tr}
\begin{figure}
    \includegraphics[width=\linewidth]{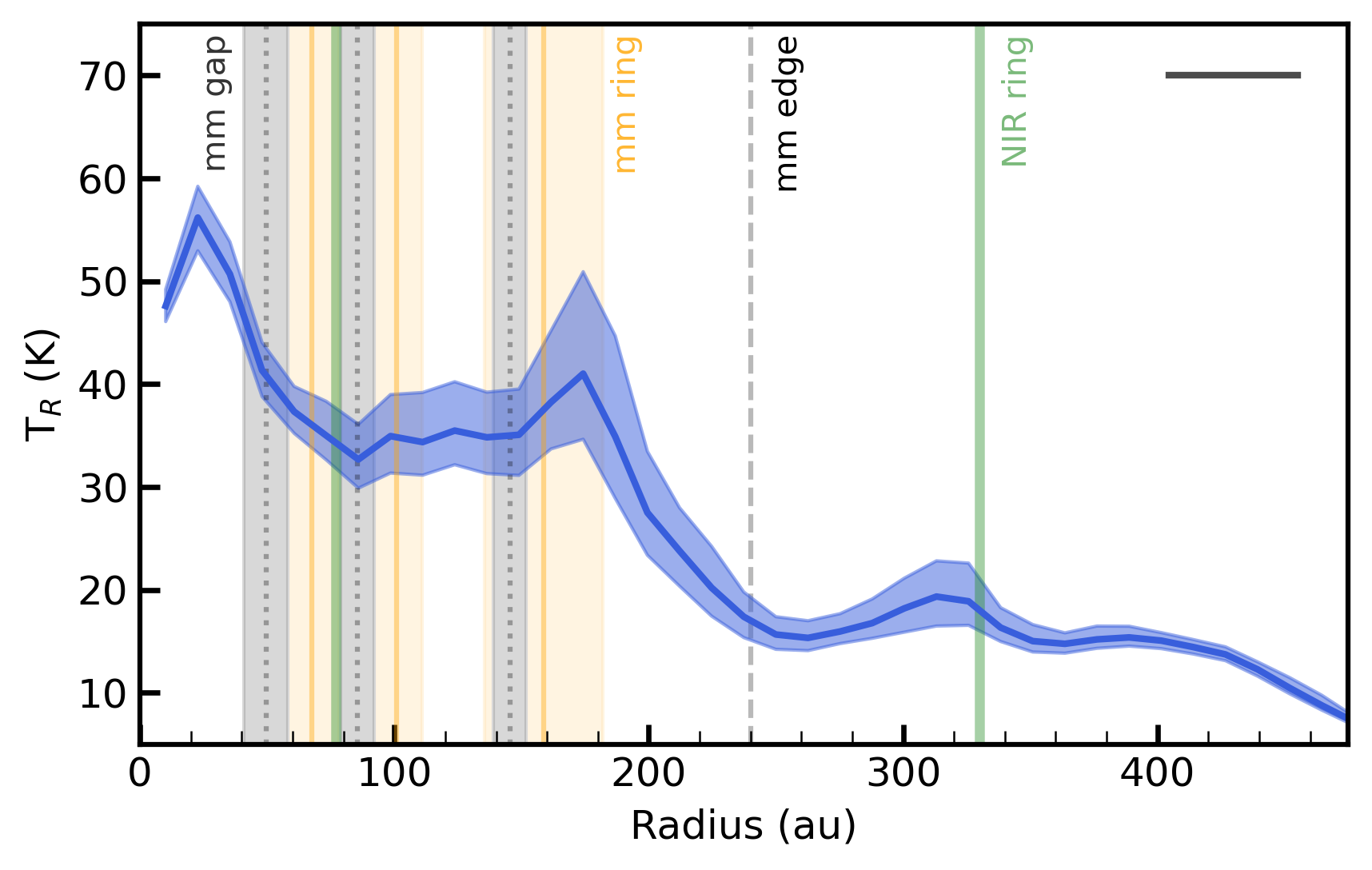}
    \caption{CN rotational temperature profile derived from joint fitting of N=1--0 and N=2--1 lines in HD 163296.  Vertical lines represent the locations of millemeter rings (solid orange), millimeter gaps (dotted grey), the millimeter disk edge (dashed grey), and NIR rings (solid green).  For millimeter features, the measured substructure widths are indicated with shaded regions.  The restoring beam FWHM is shown with a horizontal black line.}
    \label{fig:joint_temp}
\end{figure}

\begin{figure*}
    \includegraphics[width=\linewidth]{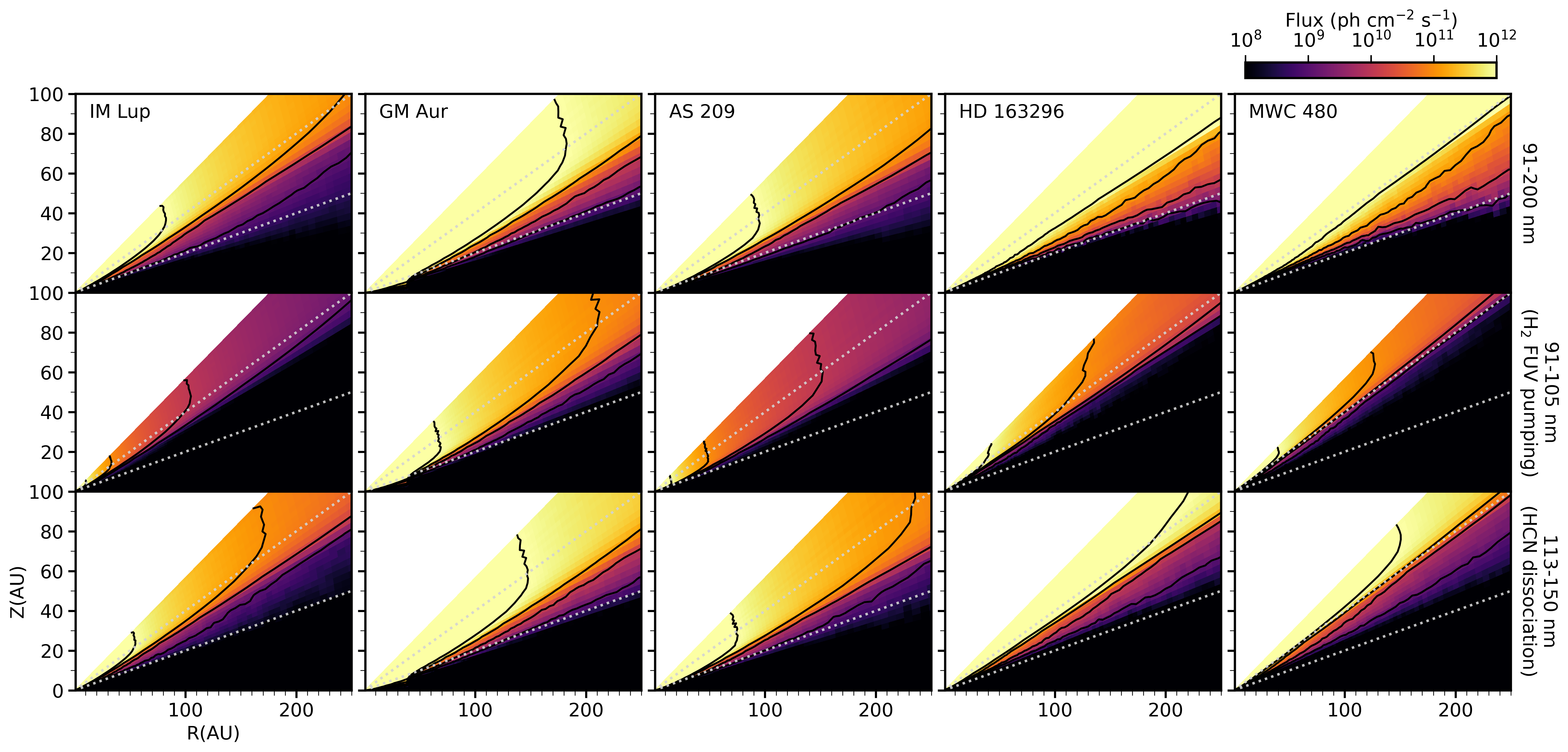}
    \caption{Toy disk modeling results showing the distribution of total UV flux (top row) as well as the flux within the wavelength bins 91-105 nm (i.e.\ CN formation via FUV pumping of H$_2$; middle row) and 113-150 nm (HCN selective photodissociation; bottom row).  Contour levels correspond to 10$^{9}$, 10$^{10}$, 10$^{11}$, and 10$^{12}$ photon cm$^{-2}$  s$^{-1}$.  Dotted grey lines mark elevations with $z/r$=0.2 and 0.4 for reference.}
    \label{fig:UV}
\end{figure*}

In the one case where we have both CN N=2--1 and N=1--0 observations, HD 163296, we obtain good constraints on the radial CN rotational temperature.  Figure \ref{fig:joint_temp} shows the CN rotational temperature profile in HD 163296, along with dust substructure locations observed at millimeter wavelengths \citep{Law2020_rad} and NIR wavelengths \citep{Monnier2017, Muro-Arena2018, Rich2020}.  The CN rotational temperatures are warm in the inner $\sim$200 au (30--60 K), and drop to 15--20 K beyond this.  We also observe several peaks in the rotational temperature, occurring around 20, 170, and 320 au.  Using the same criteria described in Section \ref{sec:line_vs_cont}, temperature maxima appear to be associated with the major dust rings at 170 au (millimeter) and 320 au (NIR).  Dust rings in the inner 100 au are not accompanied by CN rotational temperature peaks.  However, at the moderate (0.5'') spatial resolution of this data, there are just two resolving elements across this region, and higher-resolution observations are needed to determine if there are small-scale CN temperature peaks in the inner disk.  It is also important to note that if CN is emitting from a flared surface, the radial location of the CN features may be shifted with respect to the midplane pebble population, thus complicating a comparison with dust features seen at millimeter wavelengths.  Still, we see tentative evidence for increased CN rotational temperatures near the presence of dust rings, which we discuss further in Section \ref{sec:disc_CN_layer}.

\section{Toy model: Local UV fields}
\label{sec:toymodel}

\subsection{Model setup}
In Section \ref{subsec:col_uv}, we showed that stellar UV luminosities are poor predictors for the peak or average CN, HCN or CN/HCN column densities in the MAPS disks.  Here, we present a toy modeling framework to explore the local UV fluxes in the MAPS disks, which are regulated by the disk physical structure as well as the incident UV field.  With this we can identify patterns between the observed CN and HCN emission and the source-dependent propagation of UV radiation within the disk.

Source-specific disk physical models for all five MAPS sources were derived in MAPS V \citep{Zhang2020} by simultaneously fitting the SED, 1.3~mm continuum image, and photospheric, UV, and X-ray spectra.  We adopt the resulting dust density profiles, which include both a small dust population (0.005--1~$\mu$m) and large dust population (0.005~$\mu$m--1~mm).  We then use RADMC-3D \citep{Dullemond2012}, along with the same dust opacities and stellar spectra adopted in \citet{Zhang2020} and assuming isotropic scattering, to simulate the local UV field at each grid position in the disk.  

Figure \ref{fig:UV} shows the modeled UV fluxes for the five MAPS disks.  In addition to the total UV flux, we explore whether any trends exist within the wavelength regions where specific photoprocesses are activated.  Namely, CN formation via FUV pumping of H$_2$ should be efficient at wavelengths $<$105~nm, while HCN should be selectively photodissociated over CN at wavelengths $>$113~nm (Section \ref{sec:intro}).  Thus, the total UV flux as well as the flux within these wavelength ranges are shown in the top, middle, and bottom rows of Figure \ref{fig:UV}.

\subsection{FUV-induced CN formation}
We first compare the toy UV models with the observed CN emission morphology (Figures \ref{fig:CN_mom0} and \ref{fig:CN_radprofs}).  GM Aur is the only disk with centrally peaked emission, exhibiting bright CN within the inner $\sim$50 au.  This is very likely a result of the inner dust cavity in GM Aur.  As can be seen in Figure \ref{fig:UV}, the UV flux penetrates closer to the disk midplane within the dust-depleted inner cavity, and thus UV photoprocesses like CN formation should be enhanced in this region.

In all other disks, the CN emission exhibits a central depression.  Modeling by \citet{Cazzoletti2018} shows that ringed CN emission is a common outcome when CN formation is initiated by the production of vibrationally excited H$_2^*$.  In turn, H$_2^*$ is limited in the inner-most disk regions by some combination of efficient H$_2$ dissociation or collisional de-excitation of H$_2^*$.  Observations by \citet{Arulanantham2020} also show evidence for CN destruction in strong FUV fields, which could additionally contribute to the CN emission deficit in the inner disk.  Of these three mechanisms to explain a central CN depression (i.e., H$_2$ photodissociation, H$_2^*$ collisional de-excitation, or CN photodissociation), photodissociation of H$_2$ or CN is at odds with the CN emission peak observed within the inner cavity of GM Aur, since efficient dissociation in the UV-exposed cavity should produce a strong CN deficit instead of a peak.  In contrast, collisional H$_2^*$ de-excitation is consistent with both the observed CN emission deficits in the high-density inner regions of the full disks in our sample, as well as the CN emission peak in the low-density inner cavity of the transition disk GM Aur.  Thus, we speculate that collisional H$_2^*$ de-excitation plays the dominant role in shaping the observed inner CN holes in full disks.

As discussed in Section \ref{subsec:obs_morphologies}, all disks exhibit a compact emission peak at a few tens of au (though this feature is quite small in the case of IM Lup), followed by a drop in intensity beyond 50--150 au.  Here we highlight that, in Figure \ref{fig:UV}, the total UV photon flux (top row) is not scarce beyond these radii.  However, when considering only the flux of photons with wavelengths $<$105 nm (i.e. those corresponding to H$_2$ pumping; middle row), the flux is attenuated more steeply with radius.  This would provide a natural explanation for the bright CN rings that fall off rapidly in intensity beyond 50--150 au.  Moreover, we note that IM Lup, which exhibits only a weak inner CN component, is also predicted to have a low flux of $<$105 nm photons at essentially all radii.  The high modeled $<$105 nm flux in GM Aur is in some tension with the relatively weak observed CN emission.  However, the input spectrum for GM Aur is particularly uncertain as the star is known to exhibit high UV variability \citep{Robinson2019}.

\subsection{The role of selective HCN photodissociation}
\label{subsec:model_hcn}
Models also predict that the CN/HCN ratio should be sensitive to selective photodissociation of HCN \citep[e.g.][]{Aikawa1999, Fogel2011}, which is most efficient at UV wavelengths $\sim$113--150~nm.  Based on the photon fluxes in Figure \ref{fig:UV}, we would expect the highest CN/HCN ratios in HD 163926, MWC 480, and GM Aur (though, as noted above, this flux may be an over-estimate), and lower CN/HCN ratios in AS 209 and IM Lup.  From the peak and average column density ratios presented in Table \ref{tab:columns}, however, we see that IM Lup has a high CN/HCN ratio relative to the rest of the sample, and GM Aur the lowest.  HD 163296 and MWC 480 both show relatively high CN/HCN ratios, as expected in a photodissociation framework.  However, the data set as a whole seems incompatible with a simple relationship between the UV flux and bulk metrics of the CN/HCN ratio (e.g.~peak or average across the disk).   

\section{Discussion}
\label{sec:discussion}

\subsection{CN formation pathways}
\label{subsec:cn_holes}

A main motivation for the modeling study of \citet{Cazzoletti2018} was to explain the dramatic ringed morphology of CN in TW Hya, originally presented in \citet{Teague2016}.  Those models show that CN rings are a natural outcome when CN formation is initiated by vibrationally excited H$_2^*$.  This is largely due to the ringed morphology of H$_2^*$ itself: the abundance is low in the inner disk due to some combination of H$_2$ photodissociation and H$_2^*$ collisional de-excitation, and also low in the outer disk where shielding inhibits FUV pumping to form H$_2^*$.  The CN abundance profile thus inherits this ringed profile.  Subsequently, the disks Sz 71 and Sz 98 in Lupus were found to have a similar morphology consisting of a single bright CN emission ring \citep{vanTerwisga2019}, which can be readily explained by H$_2^*$-initiated formation chemistry.  

In our sample (Figure \ref{fig:CN_radprofs}), we observe a bright inner CN emission ring towards AS 209, HD 163296, and MWC 480, while IM Lup exhibits a weak emission ring, and GM Aur (a transition disk) shows a bright central peak.  Thus, in disks without a dust cavity, an emission depression in the inner tens of au does seem to be a common morphology for CN, as predicted by the models of \citet{Cazzoletti2018}.  Moreover, our toy models (Figure \ref{fig:UV}) show that the bright, compact emission features seen towards our sample are readily explained by the availability of FUV ($<$105 nm) photons which form H$_2^*$.  Thus, in general our observations support that this chemistry is important for CN formation in the inner $\sim$150 au.

However, in all five disks we also observe a weaker, more diffuse CN emission component beyond the bright inner component.  This component extends to $\sim$600 au in all disks except the relatively compact disk AS 209, in which case CN still exhibits the largest radial extent of any molecule except CO \citep{Law2020_rad}.  It is unlikely that the same H$_2^*$-driven chemistry can explain this extended CN emission component, given that observations show UV-fluorescent H$_2$ emission only from the inner disk \citep[e.g.][]{France2012, Arulanantham2020}.  A separate CN formation pathway (or multiple pathways) operating in conditions of low density and moderate UV flux (perhaps with a broader range of wavelengths than the H$_2^*$-driven chemistry) is likely needed to explain the diffuse outer emission shelf.  Further modeling work is needed to explore what chemistry is responsible for the diffuse CN component, as well as the role of external vs.\ stellar UV in driving its formation.

\subsection{CN emitting layer}
\label{sec:disc_CN_layer}
The models of \citet{Cazzoletti2018} and \citet{Visser2018} predict that CN formation should peak at altitudes of $z/r \sim$0.2--0.4.  \citet{Teague2020} similarly infer that CN emits from $z/r >$ 0.2 in the TW Hya disk.  From our toy models (Figure \ref{fig:UV}), we see that the elevation range $z/r$=0.2--0.4 coincides with the disk regions with high UV fluxes, including the FUV wavelengths required for H$_2^*$ formation.  The CN rotational temperature in the joint N=2--1 and 1--0 fitting for HD 163296 provides an additional constraint on the CN emitting layer.  The temperature profile we derived indicates that CN emits from a slightly cooler layer compared to $^{12}$CO (Figure \ref{fig:HD163296_CN_profs}).  The $^{12}$CO emission surface in the MAPS disks ranges from $z/r$=0.2--0.5 \citep{Law2020_surf}, again consistent with a CN emission surface of $z/r$=0.2--0.4.  Note, however, that the lower CN rotational temperature beyond $\sim$200 au coincides with the diffuse emission shelf seen in the column density profile.  This likely corresponds to a different formation chemistry than the H$_2^*$-driven formation in the inner disk (Section \ref{subsec:cn_holes}), and may correspondingly emit from a lower elevation.  At least in the inner $\sim$200 au, however, it appears that the CN emission surface is higher than the HCN surface, which is consistent with a $z/r \sim$0.1 \citep{Law2020_surf}.  This vertical stratification agrees with model predictions that HCN should exist in more UV-shielded disk regions compared to CN \citep[e.g.][]{Aikawa1999, vanZadelhoff2003, Fogel2011}.  

In Figure \ref{fig:joint_temp}, we showed that the CN rotational temperatures in HD 163296 appear to increase near the dust rings around 170 and 320 au.  A possible explanation for this is that the increase in UV opacity at radii with dust over-densities pushes the CN formation layer to higher (and warmer) disk regions.  From Figure \ref{fig:Nt_substructures}, the CN column density profiles do not exhibit a clear response to the presence of these rings, indicating that the chemistry proceeds with a similar efficiency in the more elevated layer.  Constraints on the CN emission surface are needed to determine if this effect is real, or if the correspondence between rotational temperature peaks and dust rings is coincidental.

\subsection{Radial trends in CN and HCN column densities}
\label{subsec:disc_radial}

It is of great interest to understand whether chemical substructures identified in protoplanetary disks are connected to the dust substructures \citep[see][]{Law2020_rad}.  In the case of CN and HCN, variations in the dust distribution would be expected to matter primarily if they cause local perturbations to the UV field.  In Section \ref{sec:line_vs_cont}, we showed that millimeter dust gaps and rings often (but not always) appear coincident with peaks and troughs, respectively, in the CN/HCN ratio (Figure \ref{fig:Nt_substructures}).

The association of multiple millimeter gaps with enhanced CN/HCN ratios in our sample is consistent with a scenario in which a diminished dust population promotes UV penetration within the gap \citep[e.g.][]{Alarcon2020}, driving the production of CN and/or the destruction of HCN.  Correspondingly, dust overdensities at millimeter ring positions should inhibit UV penetration, either suppressing CN formation or pushing it to a higher disk elevation, while inhibiting HCN photodissociation.  The fact that this association is identified for some millimeter features but is not universal could reflect the varying properties of different dust gaps and rings, translating to different impacts on the UV photochemistry.  Indeed, different gap-opening mechanisms are predicted to have varying effects on the distributions of gas and small and large dust grains \citep[e.g.][]{Dong2015,Rosotti2016, Pinilla2017}.  We also note that CN emission substructures may be unresolved at the spatial resolution used for this analysis (0.3''), and higher-resolution CN observations are needed to more quantitatively compare the CN/HCN substructure positions and contrasts with the analogous dust properties.  If the degree of CN/HCN perturbation is indeed sensitive to dust substructure characteristics such as contrast and vertical extent, then the CN/HCN ratio may ultimately be a useful tool in inferring the distribution of solids within rings and gaps.

NIR rings are not consistently associated with features in the CN/HCN ratio (Figure \ref{fig:columns_nir}).  The lack of a clear correspondence between NIR rings and the CN/HCN ratio could be related to the relatively low contrast of many features detected in the NIR.  In this case, small enhancements to the small dust population may simply push CN production to slightly higher disk layers without strongly impacting the column density.  As shown in Section \ref{sec:disc_CN_layer}, this is supported by the CN rotational temperature profile derived for HD 163296.  Good constraints on the CN rotational temperatures and emitting layers in additional sources are needed to confirm whether shifts in the CN formation layer can indeed explain the lack of clear relationship between NIR rings and the CN/HCN ratio.

In addition to the coincidence between some millimeter dust gaps and rings with CN/HCN peaks and troughs, respectively, the variation in CN/HCN ratio from the inner disk to the outer disk provides further evidence that the dust population plays an important role in regulating the CN and HCN chemistries. The radial CN/HCN ratios presented in Figure \ref{fig:cn_hcn_columns} increase by $\sim$2 orders of magnitude with increasing radius in the MAPS disks.  This is likely related to enhanced UV penetration in the lower-density, less-shielded outer disk, which will result in selective photodissociation of HCN over CN.  Indeed, an increase in the CN/HCN column density ratio in the outer disk is a common outcome of disk chemistry models \citep[e.g][]{Aikawa2002, Jonkheid2007}.  The radial CN/HCN ratio may therefore be a useful indicator of the UV penetration at a given location in the disk.

\subsection{CN and HCN trends across the disk sample}
\label{subsec:disc_disks}
Previous disk chemistry surveys by \citet{Kastner2008}, \citet{Salter2011}, and \citet{Oberg2011} have found roughly constant CN/HCN flux ratios across their disk samples.  While line ratios are subject to optical depth effects, our results using column density ratios confirm that the peak and average CN/HCN ratios are consistent within a factor of a few across our sample (Figure \ref{fig:lum_columns}).  Moreover, our derived CN/HCN ratios show no trend with stellar UV luminosity, which varies by two orders of magnitude across the sample.  We similarly found no clear relationship between the bulk disk CN/HCN ratios and the modeled UV fields (Section \ref{subsec:model_hcn}). Thus, while the CN/HCN ratio appears sensitive to changes in the physical environment within a given disk (Section \ref{subsec:disc_radial}), it is not a useful metric of the incident UV flux to the disk.

The CN column densities are similarly consistent across the disk sample, varying by just factors of a few in N$_{\mathrm{max}}$ or N$_{\mathrm{avg}}$.  We speculate that the CN column densities are similar even in disks with very different physical properties because the underlying chemistry is similar, but the regions where it is efficient differ.  Similarly, morphological differences in CN emission across the sample (i.e. the shape and radial extent of the central CN ring; Figure \ref{fig:CN_radprofs}) may reflect that the specific conditions needed for CN formation chemistry are met in different radial and vertical regions depending on the source-specific combination of stellar properties (i.e.\ luminosity) and disk properties (i.e.\ dust density structures).  A similar effect was found to explain the widely varying C$_2$H morphologies presented in \citet{Bergner2019}.  Thus, our findings are consistent with CN formation in a narrow range of physical conditions, resulting in similar column densities but some morphological variations across the disk sample.

\subsection{Chemical connection between CN and HCN}
\label{subsec:disc_corr}
We find that the local CN and HCN column densities within different radial bins trend positively within one another (Figure \ref{fig:cn_vs_hcn}).  Moreover, the column densities from different disks fall along a similar trend line, such that the local CN column density can be inferred from the HCN column density.  This indicates that the absolute efficiencies of CN and HCN formation are connected.  Meanwhile, the relative abundance of CN vs.\ HCN can vary significantly within a disk as a result of the local disk conditions, e.g.\ enhanced HCN photodissociation in UV-exposed regions.  

A common dependence on the C/O ratio in the gas could explain the positive trend between CN and HCN column densities, as nitrile formation should in general depend on the availability of free carbon \citep[e.g.][]{LeGal2019}.  \citet{Cazzoletti2018} find that CN production is only weakly sensitive to the C/O ratio, however this is consistent with the narrow range in CN column densities observed across our sample.  Disk chemical models that address CN and HCN formation simultaneously are needed to further explore how the C/O ratio (and other factors such as UV exposure) impact the absolute and relative CN and HCN chemistries.

As noted in Section \ref{sec:disc_CN_layer}, CN seems to emit from a higher vertical layer than HCN.  That CN and HCN column densities are positively correlated despite emitting from different vertical layers implies that either dynamical mixing effectively connects the different layers, or that the conditions required for nitrile formation (e.g.\ high C/O) are met across a wide vertical range within the disk.  Such a chemical connection across different disk layers is intriguing from the standpoint of understanding the compositions of planet-accreting gas.  Further work is required to explore this finding in more detail, for instance testing whether the local column densities of complex nitriles emitting closer to the midplane \citep{Ilee2020} show a similar correlation with the CN and HCN column densities.

\section{Conclusions}
\label{sec:conclusion}
As part of the MAPS program, we have analyzed CN and HCN emission towards five protoplanetary disks in order to explore the relationship between these small nitriles and the UV field.  Our conclusions are as follows:

\begin{enumerate}
    \item CN emission exhibits an inner ring for all disks without an inner dust cavity. GM Aur, a transition disk, instead shows centrally peaked emission.  Toy modeling of the local UV field indicates that the bright, compact CN emission component can be explained by the availability of FUV ($<$105 nm) photons, which are predicted to drive CN formation via the production of vibrationally excited H$_2$ \citep{Cazzoletti2018}.  The CN rotational temperatures are also consistent with an elevated CN emission surface ($z/r \sim$0.2--0.4) as predicted by this framework.

    \item All five disks also exhibit a diffuse CN emission component extending a few hundred au.  This component likely forms through a different pathway than the H$_2^*$-driven chemistry in the inner disk.
    
    \item Despite two orders of magnitude variation in the stellar UV luminosity across our disk sample, the peak and average CN/HCN ratios vary by just a factor of a few.  We conclude that the CN/HCN column density ratio is not a useful predictor of the incident UV flux from disk to disk. 
    
    \item In all disks, the CN/HCN column density ratio increases with radius from about unity to 100.  This likely reflects that HCN can be selectively photodissociated by UV photons ($\lambda >$113~nm), which will experience less shielding as dust densities decrease in the outer disk.  The CN/HCN ratio is therefore sensitive to radial differences in UV penetration within a disk.
    
    \item Many but not all millimeter dust gaps and rings are associated with, respectively, peaks and troughs in the CN/HCN column density ratio.  This is again consistent with enhanced HCN destruction in disk regions with greater UV penetration.  Dust substructure properties such as contrast and vertical extent may dictate the degree to which the CN/HCN ratio is impacted, and higher-resolution observations are needed to further explore this connection.
    
    \item Local CN and HCN column densities exhibit a positive correlation, possibly due to a common dependence on the C/O ratio.  Given that CN appears to emit from a higher elevation than HCN, this correlation implies that different disk vertical layers are chemically linked.

\end{enumerate}

Our findings have several implications for the compositions of forming planets.  First, the robust FUV-driven formation of CN in the inner $\sim$150 au in the MAPS disks is likely accompanied by the formation of other strongly photo-sensitive species.  At these disk radii, we expect planet-accreting gas from somewhat elevated disk layers to have photochemically dominated abundance patterns.  At larger radii, the increasing CN/HCN ratio indicates that UV penetration into deeper disk layers becomes more efficient.  Thus, we expect a moderate photochemistry to influence the molecular layer compositions at outer disk radii.  Additionally, given that the local CN and HCN column densities are positively correlated despite emitting from different elevations, the different disk layers appear chemically linked through either dynamics or through shared physical/chemical conditions (e.g.\ C/O ratio).  Lastly, the CN/HCN ratio is $>$1 at all disk radii in AS 209 and IM Lup, and beyond $\sim$50 au in the remaining MAPS disks, meaning that CN is the dominant carrier of the prebiotically interesting nitrile group \citep[e.g.][]{Powner2009} at most disk radii.  

Important unknowns remain in our understanding of CN and CN/HCN chemistry in disks, including the chemistry responsible for the diffuse outer disk CN emission component, the impact of C/O on the relative CN and HCN chemistries, and the relationship between millimeter dust substructure properties and the CN/HCN ratio.  An increased sample size of higher-sensitivity and higher-SNR observations (readily achievable by targeting CN N=3--2 transitions instead of N=1--0), coupled with more sophisticated disk chemistry modeling, are needed to address these questions.  

\section*{Acknowledgements} 
We are grateful to the anonymous referee for feedback on this manuscript.  This paper makes use of ALMA data, project codes 2018.1.01055.L (P.I.: K. \"Oberg) and 2016.1.0084.S (P.I: V. Guzm\'an).  ALMA is a partnership of ESO (representing its member states), NSF (USA), and NINS (Japan), together with NRC (Canada) and NSC and ASIAA (Taiwan), in cooperation with the Republic of Chile. The Joint ALMA Observatory is operated by ESO, AUI/NRAO, and NAOJ. The National Radio Astronomy Observatory is a facility of the National Science Foundation operated under cooperative agreement by Associated Universities, Inc.

J.B.B. acknowledges support from NASA through the NASA Hubble Fellowship grant \#HST-HF2-51429.001-A awarded by the Space Telescope Science Institute, which is operated by the Association of Universities for Research in Astronomy, Incorporated, under NASA contract NAS5-26555. 
K.I.\"O. acknowledges support from the Simons Foundation (SCOL \#321183) and an NSF AAG Grant (\#1907653). 
G.C. is supported by NAOJ ALMA Scientific Research Grant Code 2019-13B
V.V.G. acknowledges support from FONDECYT Iniciaci\'on 11180904. 
C.J.L. acknowledges funding from the National Science Foundation Graduate Research Fellowship under Grant DGE1745303.
R.T. acknowledges support from the Smithsonian Institution as a Submillimeter Array (SMA) Fellow.
A.D.B. and E.A.B. acknowledge support from NSF AAG Grant \#1907653.
Y.A. acknowledges support by NAOJ ALMA Scientific Research Grant Codes 2019-13B, and Grant-in-Aid for Scientific Research 18H05222 and 20H05847.
S. M. A. and J. H. acknowledge funding support from the National Aeronautics and Space Administration under Grant No. 17-XRP17 2-0012 issued through the Exoplanets Research Program.  J. H. acknowledges support for this work provided by NASA through the NASA Hubble Fellowship grant \#HST-HF2-51460.001-A awarded by the Space Telescope Science Institute, which is operated by the Association of Universities for Research in Astronomy, Inc., for NASA, under contract NAS5-26555.
A.S.B. acknowledges the studentship funded by the Science and Technology Facilities Council of the United Kingdom (STFC).
L.I.C. gratefully acknowledges support from the David and Lucille Packard Foundation and Johnson \& Johnson's WiSTEM2D Program.
I.C. was supported by NASA through the NASA Hubble Fellowship grant HST-HF2-51405.001-A awarded by the Space Telescope Science Institute, which is operated by the Association of Universities for Research in Astronomy, Inc., for NASA, under contract NAS5-26555.
F.L. acknowledges support from the Smithsonian Institution as a Submillimeter Array (SMA) Fellow.
F.M. acknowledges support from ANR of France under contract ANR-16-CE31-0013 (Planet-Forming-Disks) and ANR-15-IDEX-02 (through CDP "Origins of Life").
K.R.S. acknowledges the support of NASA through Hubble Fellowship Program grant HST-HF2-51419.001, awarded by the Space Telescope Science Institute,which is operated by the Association of Universities for Research in Astronomy, Inc., for NASA, under contract NAS5-26555.
J.D.I. acknowledges support from the Science and Technology Facilities Council of the United Kingdom (STFC) under ST/T000287/1.
R.L.G. acknowledges support from a CNES fellowship grant.
C.W. acknowledges financial support from the University of Leeds, STFC and UKRI (grant numbers ST/R000549/1, ST/T000287/1, MR/T040726/1).
H.N. acknowledges support by NAOJ ALMA Scientific Research Grant Codes 2018-10B and Grant-in-Aid for Scientific Research 18H05441.
T.T. is supported by JSPS KAKENHI Grant Numbers JP17K14244 and JP20K04017.
Y.Y. is supported by IGPEES, WINGS Program, the University of Tokyo.

\software{
\texttt{Matplotlib} \citep{Hunter2007}, 
\texttt{Numpy} \citep{vanderWalt2011},
\texttt{RADMC3D} \citep{Dullemond2012}, 
\texttt{Astropy} \citep{Astropy2013}, 
\texttt{emcee} \citep{Foreman-Mackey2013}, 
\texttt{bettermoments} \citep{Teague2018},
\texttt{gofish} \citep{Teague2019},
\texttt{Scipy} \citep{SciPy2020}
}

\FloatBarrier
\appendix 

\section{Joint CN N=1--0 and N=2--1 fitting}
\label{sec:app_columns}

\begin{figure}
    \includegraphics[width=0.5\linewidth]{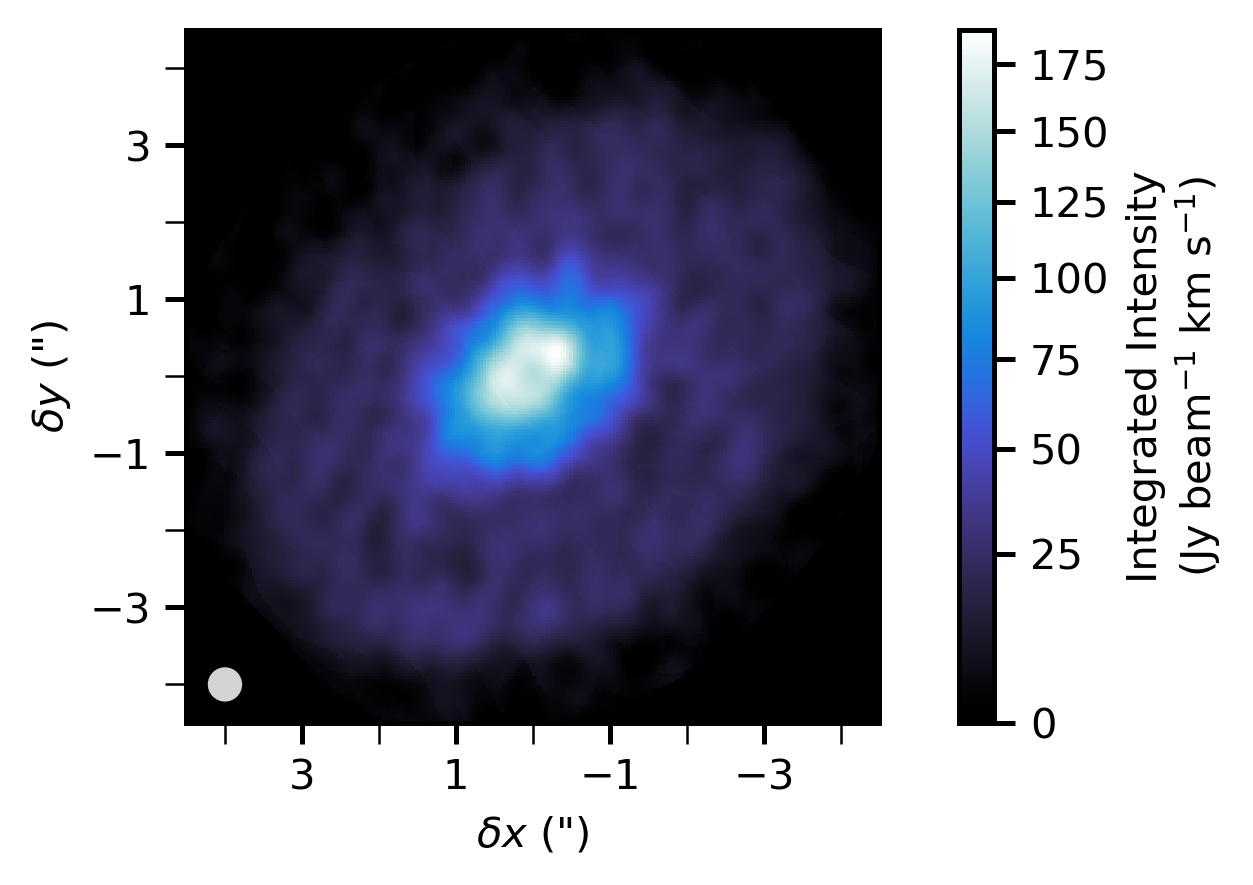}
    \caption{CN N=2-1 moment zero map in HD 163296.  To better illustrate weak emission features, integrated intensities are shown with a power-law color stretch.  The restoring beam is shown in the bottom left.}
    \label{fig:mom0_cn21}
\end{figure}

Figure \ref{fig:mom0_cn21} shows the moment zero image of the CN N=2--1 transitions at 226.659 and 226.663 GHz in HD 163296.  To perform joint spectral fitting of the CN N=2--1 and N=1--0 lines, we use a CN N=1--0 image tapered to a 0.5'' resolution to match the N=2--1 resolution.  In 0.2 km/s channels, the channel rms of this image cube is 1.4 mJy/beam.  We generated radial spectra in bins 1/4 of the beam size, and assumed a calibration uncertainty of 10\% between the two data sets.  We adopted an uninformative prior on the excitation temperature of 5 K $<$ T$_R$ $<$ 100 K.  

\begin{figure}
\centering
    \includegraphics[width=0.9\linewidth]{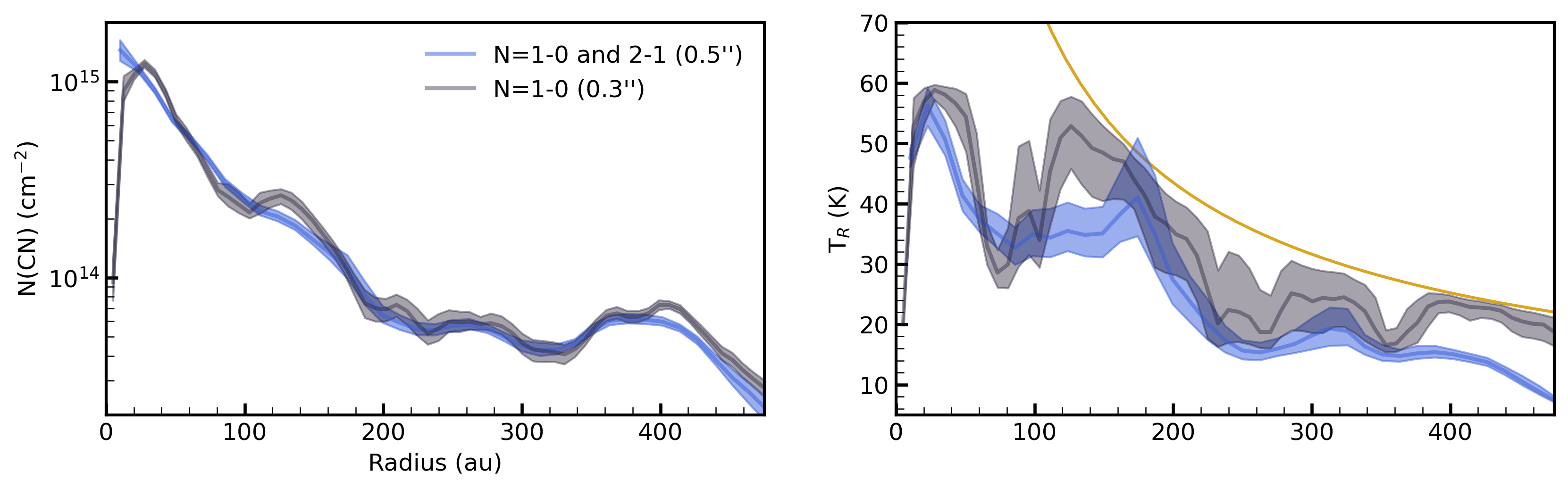}
    \caption{The column density and excitation temperature profiles in HD 163296 when the CN N=1--0 hyperfine components are fit alone (grey lines) compared to a joint fit of the N=1--0 and 2--1 lines (blue lines).  The temperature profile derived from $^{12}$CO in \citet{Law2020_surf} is shown as a gold line.}
    \label{fig:HD163296_CN_profs}
\end{figure}

Figure \ref{fig:HD163296_CN_profs} shows the resulting column density and rotational temperature profiles obtained from the joint N=1--0 and N=2--1 fitting, overlaid with the fits of the N=1--0 lines alone. For the latter, we imposed an upper bound on the rotational temperature that is the minimum between T$_{\rm{CO}}$(R) and 60 K.  As seen in Figure \ref{fig:HD163296_CN_profs}, the $^{12}$CO temperature profile derived in \citet{Law2020_surf} provides a conservative upper bound to the CN excitation temperature derived from the joint fit.  In the inner 100 au the parametric CO temperature is not constraining, and the maximum CN excitation temperature derived in the joint fit ($\sim$55 K), is used to inform the maximum allowed rotational temperature in fitting the N=1--0 lines alone.  The resulting N=1--0 column density profile is in good agreement with the joint fit, considering the differences in spatial resolution.  Importantly, the temperature profiles between the N=1--0 and joint fits do not match exactly, but the similarity in column density profiles demonstrates that our fitting is robust as long as reasonable upper limits on the rotational temperature are imposed.  We therefore apply the same priors for fitting CN N=1--0 lines in all disks (Section \ref{sec:columns}), adopting the source-specific CO temperature profiles from \citet{Law2020_surf}.  We also impose the same priors for fitting the HCN J=3--2 lines.  This is a looser constraint compared to the priors imposed in \citet{Guzman2020}, but results in very similar HCN column density profiles.

\FloatBarrier
\section{Rotational temperature and optical depth profiles}
\label{sec:app_Tr_tau}

\begin{figure}
    \includegraphics[width=\linewidth]{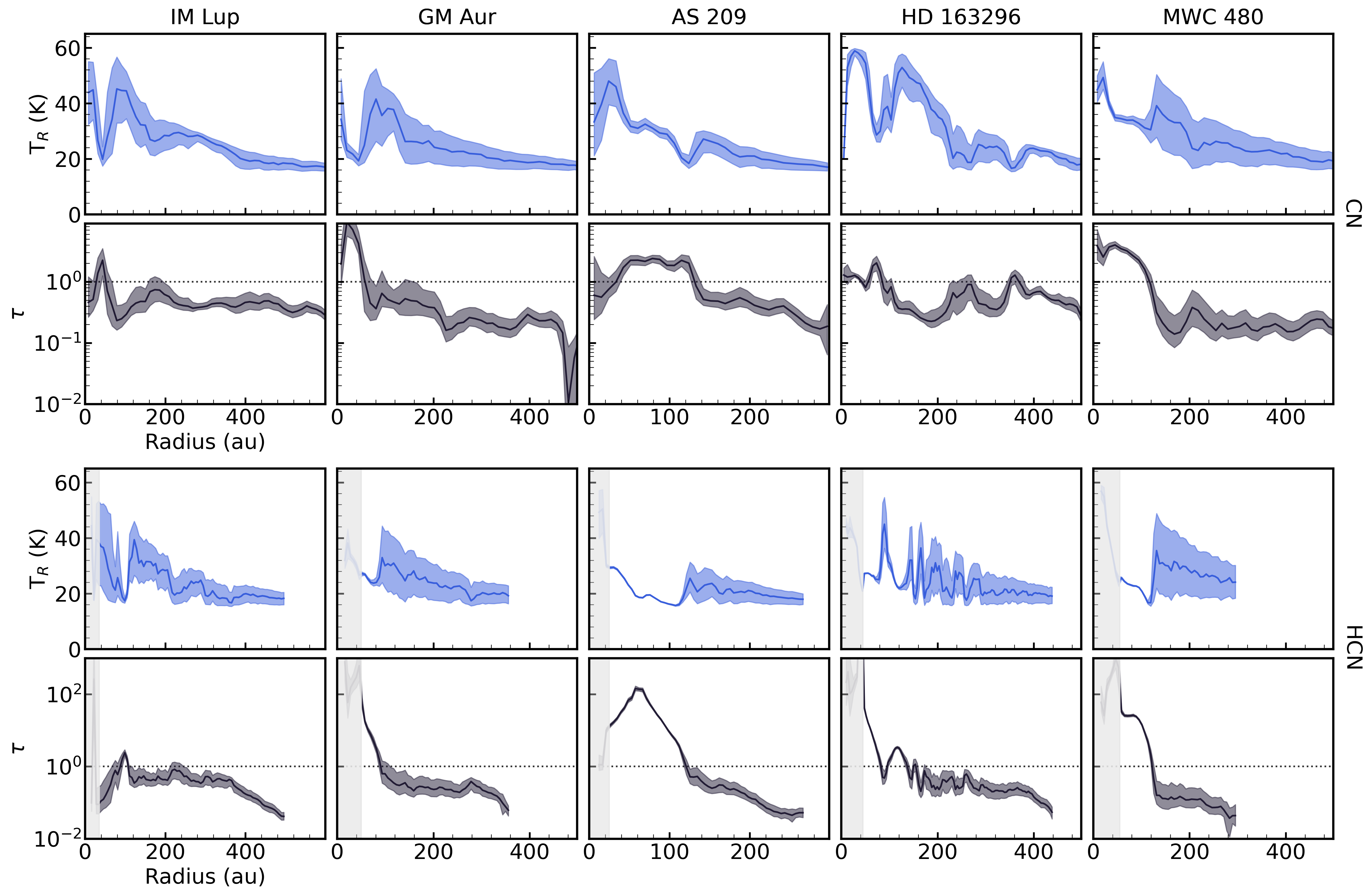}
    \caption{CN and HCN rotational temperature (blue) and optical depth (grey) profiles derived from hyperfine fitting.  $\tau$ is shown for the main hyperfine component (i.e. 113.491 GHz for CN and 265.8865 for HCN).  Shaded regions represent the 16th-84th percentiles of the fit posteriors, and solid lines represent the median.  $\tau$=1 is marked with dotted horizontal lines.}
    \label{fig:tr_tau}
\end{figure}

The rotational temperature and optical depth profiles derived from hyperfine fitting in Section \ref{sec:columns} are shown for CN and HCN in Figure \ref{fig:tr_tau}.  The rotational temperatures are generally not well constrained since all hyperfine lines of a given rotational level have the same upper state energy.  However, uncertainties in the rotational temperature do not strongly impact the resulting column density profiles (Appendix \ref{sec:app_columns}).  We find that the main CN hyperfine component (113.491 GHz) is moderately optically thick (1$< \tau < $10) at some radii but largely optically thin.  The main HCN hyperfine component (265.8865 GHz) is very optically thick ($\tau >$10) for tens of au in all disks except IM Lup.  

\FloatBarrier
\section{Comparison to NIR features}

\begin{figure}
    \includegraphics[width=0.9\linewidth]{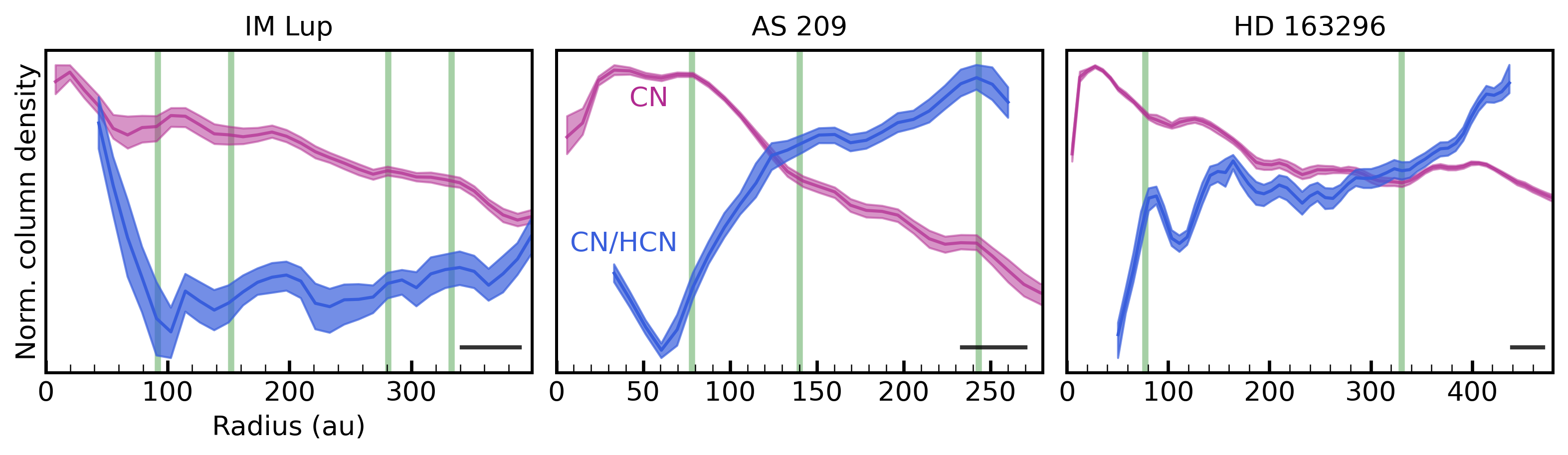}
    \begin{centering}
    \caption{Normalized column density profiles of CN and CN/HCN (pink and blue lines, respectively) with NIR ring locations overplotted (green).  NIR features are taken from \citet{Monnier2017}, \citet{Avenhaus2018}, and \citet{Rich2020}.  Horizontal black lines show the restoring beam FWHM of the line data.}
    \label{fig:columns_nir}
    \end{centering}
\end{figure}

Figure \ref{fig:columns_nir} shows the CN and CN/HCN column density profiles with NIR ring locations overplotted, for the three MAPS disks where NIR features have been reported.

\FloatBarrier

\bibliography{references}

\begin{thebibliography}{}
\expandafter\ifx\csname natexlab\endcsname\relax\def\natexlab#1{#1}\fi
\providecommand{\url}[1]{\href{#1}{#1}}
\providecommand{\dodoi}[1]{doi:~\href{http://doi.org/#1}{\nolinkurl{#1}}}
\providecommand{\doeprint}[1]{\href{http://ascl.net/#1}{\nolinkurl{http://ascl.net/#1}}}
\providecommand{\doarXiv}[1]{\href{https://arxiv.org/abs/#1}{\nolinkurl{https://arxiv.org/abs/#1}}}

\bibitem[{{Ahrens} {et~al.}(2002){Ahrens}, {Lewen}, {Takano}, {Winnewisser},
  {Urban}, {Negirev}, \& {Koroliev}}]{Ahrens2002}
{Ahrens}, V., {Lewen}, F., {Takano}, S., {et~al.} 2002, Zeitschrift
  Naturforschung Teil A, 57, 669, \dodoi{10.1515/zna-2002-0806}

\bibitem[{{Aikawa} \& {Herbst}(1999)}]{Aikawa1999}
{Aikawa}, Y., \& {Herbst}, E. 1999, \aap, 351, 233

\bibitem[{{Aikawa} {et~al.}(2002){Aikawa}, {van Zadelhoff}, {van Dishoeck}, \&
  {Herbst}}]{Aikawa2002}
{Aikawa}, Y., {van Zadelhoff}, G.~J., {van Dishoeck}, E.~F., \& {Herbst}, E.
  2002, \aap, 386, 622, \dodoi{10.1051/0004-6361:20020037}

\bibitem[{{Alarc{\'o}n} {et~al.}(2020){Alarc{\'o}n}, {Teague}, {Zhang},
  {Bergin}, \& {Barraza-Alfaro}}]{Alarcon2020}
{Alarc{\'o}n}, F., {Teague}, R., {Zhang}, K., {Bergin}, E.~A., \&
  {Barraza-Alfaro}, M. 2020, \apj, 905, 68, \dodoi{10.3847/1538-4357/abc1d6}

\bibitem[{{Andrews} {et~al.}(2018){Andrews}, {Huang}, {P{\'e}rez}, {Isella},
  {Dullemond}, {Kurtovic}, {Guzm{\'a}n}, {Carpenter}, {Wilner}, {Zhang}, {Zhu},
  {Birnstiel}, {Bai}, {Benisty}, {Hughes}, {{\"O}berg}, \&
  {Ricci}}]{Andrews2018}
{Andrews}, S.~M., {Huang}, J., {P{\'e}rez}, L.~M., {et~al.} 2018, \apjl, 869,
  L41, \dodoi{10.3847/2041-8213/aaf741}

\bibitem[{{Arulanantham} {et~al.}(2020){Arulanantham}, {France}, {Cazzoletti},
  {Miotello}, {Manara}, {Schneider}, {Hoadley}, {van Dishoeck}, \&
  {G{\"u}nther}}]{Arulanantham2020}
{Arulanantham}, N., {France}, K., {Cazzoletti}, P., {et~al.} 2020, \aj, 159,
  168, \dodoi{10.3847/1538-3881/ab789a}

\bibitem[{{Astropy Collaboration} {et~al.}(2013){Astropy Collaboration},
  {Robitaille}, {Tollerud}, {Greenfield}, {Droettboom}, {Bray}, {Aldcroft},
  {Davis}, {Ginsburg}, {Price-Whelan}, {Kerzendorf}, {Conley}, {Crighton},
  {Barbary}, {Muna}, {Ferguson}, {Grollier}, {Parikh}, {Nair}, {Unther},
  {Deil}, {Woillez}, {Conseil}, {Kramer}, {Turner}, {Singer}, {Fox}, {Weaver},
  {Zabalza}, {Edwards}, {Azalee Bostroem}, {Burke}, {Casey}, {Crawford},
  {Dencheva}, {Ely}, {Jenness}, {Labrie}, {Lim}, {Pierfederici}, {Pontzen},
  {Ptak}, {Refsdal}, {Servillat}, \& {Streicher}}]{Astropy2013}
{Astropy Collaboration}, {Robitaille}, T.~P., {Tollerud}, E.~J., {et~al.} 2013,
  \aap, 558, A33, \dodoi{10.1051/0004-6361/201322068}

\bibitem[{{Avenhaus} {et~al.}(2018){Avenhaus}, {Quanz}, {Garufi}, {Perez},
  {Casassus}, {Pinte}, {Bertrang}, {Caceres}, {Benisty}, \&
  {Dominik}}]{Avenhaus2018}
{Avenhaus}, H., {Quanz}, S.~P., {Garufi}, A., {et~al.} 2018, \apj, 863, 44,
  \dodoi{10.3847/1538-4357/aab846}

\bibitem[{{Bergin} {et~al.}(2003){Bergin}, {Calvet}, {D'Alessio}, \&
  {Herczeg}}]{Bergin2003}
{Bergin}, E., {Calvet}, N., {D'Alessio}, P., \& {Herczeg}, G.~J. 2003, \apjl,
  591, L159, \dodoi{10.1086/377148}

\bibitem[{{Bergner} {et~al.}(2019){Bergner}, {{\"O}berg}, {Bergin}, {Loomis},
  {Pegues}, \& {Qi}}]{Bergner2019}
{Bergner}, J.~B., {{\"O}berg}, K.~I., {Bergin}, E.~A., {et~al.} 2019, \apj,
  876, 25, \dodoi{10.3847/1538-4357/ab141e}

\bibitem[{{Cataldi} {et~al.}(2021){Cataldi}, {Yamato}, {Aikawa}, {Bergner},
  {Furuya}, {Guzm{\'a}n}, {Huang}, {Loomis}, {Qi}, {Andrews}, {Bergin},
  {Booth}, {Bosman}, {Cleeves}, {Czekala}, {Ilee}, {Law}, {Le Gal}, {Liu},
  {Long}, {M{\'e}nard}, {Nomura}, {{\"O}berg}, {Schwarz}, {Teague},
  {Tsukagoshi}, {Walsh}, {Wilner}, \& {Zhang}}]{Cataldi2020}
{Cataldi}, G., {Yamato}, Y., {Aikawa}, Y., {et~al.} 2021, arXiv e-prints,
  arXiv:2109.06462.
\newblock \doarXiv{2109.06462}

\bibitem[{{Cazzoletti} {et~al.}(2018){Cazzoletti}, {van Dishoeck}, {Visser},
  {Facchini}, \& {Bruderer}}]{Cazzoletti2018}
{Cazzoletti}, P., {van Dishoeck}, E.~F., {Visser}, R., {Facchini}, S., \&
  {Bruderer}, S. 2018, \aap, 609, A93, \dodoi{10.1051/0004-6361/201731457}

\bibitem[{{Chapillon} {et~al.}(2012){Chapillon}, {Guilloteau}, {Dutrey},
  {Pi{\'e}tu}, \& {Gu{\'e}lin}}]{Chapillon2012}
{Chapillon}, E., {Guilloteau}, S., {Dutrey}, A., {Pi{\'e}tu}, V., \&
  {Gu{\'e}lin}, M. 2012, \aap, 537, A60, \dodoi{10.1051/0004-6361/201116762}

\bibitem[{{Cleeves} {et~al.}(2013){Cleeves}, {Adams}, \&
  {Bergin}}]{Cleeves2013}
{Cleeves}, L.~I., {Adams}, F.~C., \& {Bergin}, E.~A. 2013, \apj, 772, 5,
  \dodoi{10.1088/0004-637X/772/1/5}

\bibitem[{{Cox} {et~al.}(1992){Cox}, {Omont}, {Huggins}, {Bachiller}, \&
  {Forveille}}]{Cox1992}
{Cox}, P., {Omont}, A., {Huggins}, P.~J., {Bachiller}, R., \& {Forveille}, T.
  1992, \aap, 266, 420

\bibitem[{{Czekala} {et~al.}(2021){Czekala}, {Loomis}, {Teague}, {Booth},
  {Huang}, {Cataldi}, {Ilee}, {Law}, {Walsh}, {Bosman}, {Guzm{\'a}n}, {Le Gal},
  {{\"O}berg}, {Yamato}, {Aikawa}, {Andrews}, {Bae}, {Bergin}, {Bergner},
  {Cleeves}, {Kurtovic}, {M{\'e}nard}, {Nomura}, {P{\'e}rez}, {Qi}, {Schwarz},
  {Tsukagoshi}, {Waggoner}, {Wilner}, \& {Zhang}}]{Czekala2020}
{Czekala}, I., {Loomis}, R.~A., {Teague}, R., {et~al.} 2021, arXiv e-prints,
  arXiv:2109.06188.
\newblock \doarXiv{2109.06188}

\bibitem[{{Dionatos} {et~al.}(2019){Dionatos}, {Woitke}, {G{\"u}del},
  {Degroote}, {Liebhart}, {Anthonioz}, {Antonellini}, {Baldovin-Saavedra},
  {Carmona}, {Dominik}, {Greaves}, {Ilee}, {Kamp}, {M{\'e}nard}, {Min},
  {Pinte}, {Rab}, {Rigon}, {Thi}, \& {Waters}}]{Dionatos2019}
{Dionatos}, O., {Woitke}, P., {G{\"u}del}, M., {et~al.} 2019, \aap, 625, A66,
  \dodoi{10.1051/0004-6361/201832860}

\bibitem[{{Dixon} \& {Woods}(1977)}]{Dixon1977}
{Dixon}, T.~A., \& {Woods}, R.~C. 1977, \jcp, 67, 3956,
  \dodoi{10.1063/1.435412}

\bibitem[{{Dong} {et~al.}(2015){Dong}, {Zhu}, \& {Whitney}}]{Dong2015}
{Dong}, R., {Zhu}, Z., \& {Whitney}, B. 2015, \apj, 809, 93,
  \dodoi{10.1088/0004-637X/809/1/93}

\bibitem[{{Dullemond} {et~al.}(2012){Dullemond}, {Juhasz}, {Pohl}, {Sereshti},
  {Shetty}, {Peters}, {Commercon}, \& {Flock}}]{Dullemond2012}
{Dullemond}, C.~P., {Juhasz}, A., {Pohl}, A., {et~al.} 2012, {RADMC-3D: A
  multi-purpose radiative transfer tool}.
\newblock \doeprint{1202.015}

\bibitem[{{Dutrey} {et~al.}(1997){Dutrey}, {Guilloteau}, \&
  {Guelin}}]{Dutrey1997}
{Dutrey}, A., {Guilloteau}, S., \& {Guelin}, M. 1997, \aap, 317, L55

\bibitem[{{Flaherty} {et~al.}(2020){Flaherty}, {Hughes}, {Simon}, {Qi}, {Bai},
  {Bulatek}, {Andrews}, {Wilner}, \& {K{\'o}sp{\'a}l}}]{Flaherty2020}
{Flaherty}, K., {Hughes}, A.~M., {Simon}, J.~B., {et~al.} 2020, \apj, 895, 109,
  \dodoi{10.3847/1538-4357/ab8cc5}

\bibitem[{{Fogel} {et~al.}(2011){Fogel}, {Bethell}, {Bergin}, {Calvet}, \&
  {Semenov}}]{Fogel2011}
{Fogel}, J. K.~J., {Bethell}, T.~J., {Bergin}, E.~A., {Calvet}, N., \&
  {Semenov}, D. 2011, \apj, 726, 29, \dodoi{10.1088/0004-637X/726/1/29}

\bibitem[{{Foreman-Mackey} {et~al.}(2013){Foreman-Mackey}, {Hogg}, {Lang}, \&
  {Goodman}}]{Foreman-Mackey2013}
{Foreman-Mackey}, D., {Hogg}, D.~W., {Lang}, D., \& {Goodman}, J. 2013, \pasp,
  125, 306, \dodoi{10.1086/670067}

\bibitem[{{France} {et~al.}(2012){France}, {Schindhelm}, {Herczeg}, {Brown},
  {Abgrall}, {Alexander}, {Bergin}, {Brown}, {Linsky}, {Roueff}, \&
  {Yang}}]{France2012}
{France}, K., {Schindhelm}, E., {Herczeg}, G.~J., {et~al.} 2012, \apj, 756,
  171, \dodoi{10.1088/0004-637X/756/2/171}

\bibitem[{{Fuente} {et~al.}(1993){Fuente}, {Martin-Pintado}, {Cernicharo}, \&
  {Bachiller}}]{Fuente1993}
{Fuente}, A., {Martin-Pintado}, J., {Cernicharo}, J., \& {Bachiller}, R. 1993,
  \aap, 276, 473

\bibitem[{{Fuente} {et~al.}(1995){Fuente}, {Martin-Pintado}, \&
  {Gaume}}]{Fuente1995}
{Fuente}, A., {Martin-Pintado}, J., \& {Gaume}, R. 1995, \apjl, 442, L33,
  \dodoi{10.1086/187809}

\bibitem[{{Guzm{\'a}n} {et~al.}(2015){Guzm{\'a}n}, {{\"O}berg}, {Loomis}, \&
  {Qi}}]{Guzman2015}
{Guzm{\'a}n}, V.~V., {{\"O}berg}, K.~I., {Loomis}, R., \& {Qi}, C. 2015, \apj,
  814, 53, \dodoi{10.1088/0004-637X/814/1/53}

\bibitem[{{Guzm{\'a}n} {et~al.}(2021){Guzm{\'a}n}, {Bergner}, {Law}, {Oberg},
  {Walsh}, {Cataldi}, {Aikawa}, {Bergin}, {Czekala}, {Huang}, {Andrews},
  {Loomis}, {Zhang}, {Le Gal}, {Alarc{\'o}n}, {Ilee}, {Teague}, {Cleeves},
  {Wilner}, {Long}, {Schwarz}, {Bosman}, {P{\'e}rez}, {M{\'e}nard}, \&
  {Liu}}]{Guzman2020}
{Guzm{\'a}n}, V.~V., {Bergner}, J.~B., {Law}, C.~J., {et~al.} 2021, arXiv
  e-prints, arXiv:2109.06391.
\newblock \doarXiv{2109.06391}

\bibitem[{{Heays} {et~al.}(2017){Heays}, {Bosman}, \& {van
  Dishoeck}}]{Heays2017}
{Heays}, A.~N., {Bosman}, A.~D., \& {van Dishoeck}, E.~F. 2017, \aap, 602,
  A105, \dodoi{10.1051/0004-6361/201628742}

\bibitem[{{Huang} {et~al.}(2018){Huang}, {Andrews}, {Dullemond}, {Isella},
  {P{\'e}rez}, {Guzm{\'a}n}, {{\"O}berg}, {Zhu}, {Zhang}, {Bai}, {Benisty},
  {Birnstiel}, {Carpenter}, {Hughes}, {Ricci}, {Weaver}, \&
  {Wilner}}]{Huang2018}
{Huang}, J., {Andrews}, S.~M., {Dullemond}, C.~P., {et~al.} 2018, \apjl, 869,
  L42, \dodoi{10.3847/2041-8213/aaf740}

\bibitem[{{Huang} {et~al.}(2020){Huang}, {Andrews}, {Dullemond}, {{\"O}berg},
  {Qi}, {Zhu}, {Birnstiel}, {Carpenter}, {Isella}, {Mac{\'\i}as}, {McClure},
  {P{\'e}rez}, {Teague}, {Wilner}, \& {Zhang}}]{Huang2020}
---. 2020, \apj, 891, 48, \dodoi{10.3847/1538-4357/ab711e}

\bibitem[{{Hunter}(2007)}]{Hunter2007}
{Hunter}, J.~D. 2007, Computing in Science and Engineering, 9, 90,
  \dodoi{10.1109/MCSE.2007.55}

\bibitem[{{Ilee} {et~al.}(2021){Ilee}, {Walsh}, {Booth}, {Aikawa}, {Andrews},
  {Bae}, {Bergin}, {Bergner}, {Bosman}, {Cataldi}, {Cleeves}, {Czekala},
  {Guzm{\'a}n}, {Huang}, {Law}, {Le Gal}, {Loomis}, {M{\'e}nard}, {Nomura},
  {{\"O}berg}, {Qi}, {Schwarz}, {Teague}, {Tsukagoshi}, {Wilner}, {Yamato}, \&
  {Zhang}}]{Ilee2020}
{Ilee}, J.~D., {Walsh}, C., {Booth}, A.~S., {et~al.} 2021, arXiv e-prints,
  arXiv:2109.06319.
\newblock \doarXiv{2109.06319}

\bibitem[{{Jonkheid} {et~al.}(2007){Jonkheid}, {Dullemond}, {Hogerheijde}, \&
  {van Dishoeck}}]{Jonkheid2007}
{Jonkheid}, B., {Dullemond}, C.~P., {Hogerheijde}, M.~R., \& {van Dishoeck},
  E.~F. 2007, \aap, 463, 203, \dodoi{10.1051/0004-6361:20065668}

\bibitem[{{Jorsater} \& {van Moorsel}(1995)}]{Jorsater1995}
{Jorsater}, S., \& {van Moorsel}, G.~A. 1995, \aj, 110, 2037,
  \dodoi{10.1086/117668}

\bibitem[{{Kastner} {et~al.}(2008){Kastner}, {Zuckerman}, {Hily-Blant}, \&
  {Forveille}}]{Kastner2008}
{Kastner}, J.~H., {Zuckerman}, B., {Hily-Blant}, P., \& {Forveille}, T. 2008,
  \aap, 492, 469, \dodoi{10.1051/0004-6361:200810815}

\bibitem[{{Kastner} {et~al.}(1997){Kastner}, {Zuckerman}, {Weintraub}, \&
  {Forveille}}]{Kastner1997}
{Kastner}, J.~H., {Zuckerman}, B., {Weintraub}, D.~A., \& {Forveille}, T. 1997,
  Science, 277, 67, \dodoi{10.1126/science.277.5322.67}

\bibitem[{{Law} {et~al.}(2021{\natexlab{a}}){Law}, {Loomis}, {Teague},
  {{\"O}berg}, {Czekala}, {Andrews}, {Huang}, {Aikawa}, {Alarc{\'o}n}, {Bae},
  {Bergin}, {Bergner}, {Boehler}, {Booth}, {Bosman}, {Calahan}, {Cataldi},
  {Cleeves}, {Furuya}, {Guzm{\'a}n}, {Ilee}, {Le Gal}, {Liu}, {Long},
  {M{\'e}nard}, {Nomura}, {Qi}, {Schwarz}, {Sierra}, {Tsukagoshi}, {Yamato},
  {van't Hoff}, {Walsh}, {Wilner}, \& {Zhang}}]{Law2020_rad}
{Law}, C.~J., {Loomis}, R.~A., {Teague}, R., {et~al.} 2021{\natexlab{a}}, arXiv
  e-prints, arXiv:2109.06210.
\newblock \doarXiv{2109.06210}

\bibitem[{{Law} {et~al.}(2021{\natexlab{b}}){Law}, {Teague}, {Loomis}, {Bae},
  {{\"O}berg}, {Czekala}, {Andrews}, {Aikawa}, {Alarc{\'o}n}, {Bergin},
  {Bergner}, {Booth}, {Bosman}, {Calahan}, {Cataldi}, {Cleeves}, {Furuya},
  {Guzm{\'a}n}, {Huang}, {Ilee}, {Le Gal}, {Liu}, {Long}, {M{\'e}nard},
  {Nomura}, {P{\'e}rez}, {Qi}, {Schwarz}, {Soto}, {Tsukagoshi}, {Yamato},
  {van't Hoff}, {Walsh}, {Wilner}, \& {Zhang}}]{Law2020_surf}
{Law}, C.~J., {Teague}, R., {Loomis}, R.~A., {et~al.} 2021{\natexlab{b}}, arXiv
  e-prints, arXiv:2109.06217.
\newblock \doarXiv{2109.06217}

\bibitem[{{Le Gal} {et~al.}(2019){Le Gal}, {Brady}, {{\"O}berg}, {Roueff}, \&
  {Le Petit}}]{LeGal2019}
{Le Gal}, R., {Brady}, M.~T., {{\"O}berg}, K.~I., {Roueff}, E., \& {Le Petit},
  F. 2019, \apj, 886, 86, \dodoi{10.3847/1538-4357/ab4ad9}

\bibitem[{{Long} {et~al.}(2018){Long}, {Pinilla}, {Herczeg}, {Harsono},
  {Dipierro}, {Pascucci}, {Hendler}, {Tazzari}, {Ragusa}, {Salyk}, {Edwards},
  {Lodato}, {van de Plas}, {Johnstone}, {Liu}, {Boehler}, {Cabrit}, {Manara},
  {Menard}, {Mulders}, {Nisini}, {Fischer}, {Rigliaco}, {Banzatti}, {Avenhaus},
  \& {Gully-Santiago}}]{Long2018}
{Long}, F., {Pinilla}, P., {Herczeg}, G.~J., {et~al.} 2018, \apj, 869, 17,
  \dodoi{10.3847/1538-4357/aae8e1}

\bibitem[{{Monnier} {et~al.}(2017){Monnier}, {Harries}, {Aarnio}, {Adams},
  {Andrews}, {Calvet}, {Espaillat}, {Hartmann}, {Hinkley}, {Kraus}, {McClure},
  {Oppenheimer}, {Perrin}, \& {Wilner}}]{Monnier2017}
{Monnier}, J.~D., {Harries}, T.~J., {Aarnio}, A., {et~al.} 2017, \apj, 838, 20,
  \dodoi{10.3847/1538-4357/aa6248}

\bibitem[{{Muro-Arena} {et~al.}(2018){Muro-Arena}, {Dominik}, {Waters}, {Min},
  {Klarmann}, {Ginski}, {Isella}, {Benisty}, {Pohl}, {Garufi}, {Hagelberg},
  {Langlois}, {Menard}, {Pinte}, {Sezestre}, {van der Plas}, {Villenave},
  {Delboulb{\'e}}, {Magnard}, {M{\"o}ller-Nilsson}, {Pragt}, {Rabou}, \&
  {Roelfsema}}]{Muro-Arena2018}
{Muro-Arena}, G.~A., {Dominik}, C., {Waters}, L.~B.~F.~M., {et~al.} 2018, \aap,
  614, A24, \dodoi{10.1051/0004-6361/201732299}

\bibitem[{{{\"O}berg} {et~al.}(2010){{\"O}berg}, {Qi}, {Fogel}, {Bergin},
  {Andrews}, {Espaillat}, {van Kempen}, {Wilner}, \& {Pascucci}}]{Oberg2010}
{{\"O}berg}, K.~I., {Qi}, C., {Fogel}, J. K.~J., {et~al.} 2010, \apj, 720, 480,
  \dodoi{10.1088/0004-637X/720/1/480}

\bibitem[{{{\"O}berg} {et~al.}(2011){{\"O}berg}, {Qi}, {Fogel}, {Bergin},
  {Andrews}, {Espaillat}, {Wilner}, {Pascucci}, \& {Kastner}}]{Oberg2011}
---. 2011, \apj, 734, 98, \dodoi{10.1088/0004-637X/734/2/98}

\bibitem[{{Oberg} {et~al.}(2021){Oberg}, {Guzman}, {Walsh}, {Aikawa}, {Bergin},
  {Law}, {Loomis}, {Alarcon}, {Andrews}, {Bae}, {Bergner}, {Boehler}, {Booth},
  {Bosman}, {Calahan}, {Cataldi}, {Cleeves}, {Czekala}, {Furuya}, {Huang},
  {Ilee}, {Kurtovic}, {Le Gal}, {Liu}, {Long}, {Menard}, {Nomura}, {Perez},
  {Qi}, {Schwarz}, {Sierra}, {Teague}, {Tsukagoshi}, {Yamato}, {van 't Hoff},
  {Waggoner}, {Wilner}, \& {Zhang}}]{Oberg2020}
{Oberg}, K.~I., {Guzman}, V.~V., {Walsh}, C., {et~al.} 2021, arXiv e-prints,
  arXiv:2109.06268.
\newblock \doarXiv{2109.06268}

\bibitem[{{Pinilla} {et~al.}(2017){Pinilla}, {Pohl}, {Stammler}, \&
  {Birnstiel}}]{Pinilla2017}
{Pinilla}, P., {Pohl}, A., {Stammler}, S.~M., \& {Birnstiel}, T. 2017, \apj,
  845, 68, \dodoi{10.3847/1538-4357/aa7edb}

\bibitem[{{Powner} {et~al.}(2009){Powner}, {Gerland}, \&
  {Sutherland}}]{Powner2009}
{Powner}, M.~W., {Gerland}, B., \& {Sutherland}, J.~D. 2009, \nat, 459, 239,
  \dodoi{10.1038/nature08013}

\bibitem[{{Rich} {et~al.}(2020){Rich}, {Wisniewski}, {Sitko}, {Grady}, {Tobin},
  \& {Fukagawa}}]{Rich2020}
{Rich}, E.~A., {Wisniewski}, J.~P., {Sitko}, M.~L., {et~al.} 2020, \apj, 902,
  4, \dodoi{10.3847/1538-4357/abb2a3}

\bibitem[{{Robinson} \& {Espaillat}(2019)}]{Robinson2019}
{Robinson}, C.~E., \& {Espaillat}, C.~C. 2019, \apj, 874, 129,
  \dodoi{10.3847/1538-4357/ab0d8d}

\bibitem[{{Rosotti} {et~al.}(2016){Rosotti}, {Juhasz}, {Booth}, \&
  {Clarke}}]{Rosotti2016}
{Rosotti}, G.~P., {Juhasz}, A., {Booth}, R.~A., \& {Clarke}, C.~J. 2016,
  \mnras, 459, 2790, \dodoi{10.1093/mnras/stw691}

\bibitem[{{Ru{\'\i}z-Rodr{\'\i}guez} {et~al.}(2021){Ru{\'\i}z-Rodr{\'\i}guez},
  {Kastner}, {Hily-Blant}, \& {Forveille}}]{Ruiz2021}
{Ru{\'\i}z-Rodr{\'\i}guez}, D., {Kastner}, J., {Hily-Blant}, P., \&
  {Forveille}, T. 2021, \aap, 646, A59, \dodoi{10.1051/0004-6361/202038209}

\bibitem[{{Salter} {et~al.}(2011){Salter}, {Hogerheijde}, {van der Burg},
  {Kristensen}, \& {Brinch}}]{Salter2011}
{Salter}, D.~M., {Hogerheijde}, M.~R., {van der Burg}, R.~F.~J., {Kristensen},
  L.~E., \& {Brinch}, C. 2011, \aap, 536, A80,
  \dodoi{10.1051/0004-6361/201015411}

\bibitem[{{Schindhelm} {et~al.}(2012){Schindhelm}, {France}, {Herczeg},
  {Bergin}, {Yang}, {Brown}, {Brown}, {Linsky}, \& {Valenti}}]{Schindhelm2012}
{Schindhelm}, E., {France}, K., {Herczeg}, G.~J., {et~al.} 2012, \apjl, 756,
  L23, \dodoi{10.1088/2041-8205/756/1/L23}

\bibitem[{{Sierra} {et~al.}(2021){Sierra}, {P{\'e}rez}, {Zhang}, {Law},
  {Guzm{\'a}n}, {Qi}, {Bosman}, {{\"O}berg}, {Andrews}, {Long}, {Teague},
  {Booth}, {Walsh}, {Wilner}, {M{\'e}nard}, {Cataldi}, {Czekala}, {Bae},
  {Huang}, {Bergner}, {Ilee}, {Benisty}, {Le Gal}, {Loomis}, {Tsukagoshi},
  {Liu}, {Yamato}, \& {Aikawa}}]{Sierra2020}
{Sierra}, A., {P{\'e}rez}, L.~M., {Zhang}, K., {et~al.} 2021, arXiv e-prints,
  arXiv:2109.06433.
\newblock \doarXiv{2109.06433}

\bibitem[{{Skatrud} {et~al.}(1983){Skatrud}, {De Lucia}, {Blake}, \&
  {Sastry}}]{Skatrud1983}
{Skatrud}, D.~D., {De Lucia}, F.~C., {Blake}, G.~A., \& {Sastry}, K.~V.~L.~N.
  1983, Journal of Molecular Spectroscopy, 99, 35,
  \dodoi{10.1016/0022-2852(83)90290-4}

\bibitem[{{Stecher} \& {Williams}(1972)}]{Stecher1972}
{Stecher}, T.~P., \& {Williams}, D.~A. 1972, \apjl, 177, L141,
  \dodoi{10.1086/181069}

\bibitem[{{Teague}(2019)}]{Teague2019}
{Teague}, R. 2019, The Journal of Open Source Software, 4, 1632,
  \dodoi{10.21105/joss.01632}

\bibitem[{{Teague} \& {Foreman-Mackey}(2018)}]{Teague2018}
{Teague}, R., \& {Foreman-Mackey}, D. 2018, {Bettermoments: A Robust Method To
  Measure Line Centroids}, v1.0,  Zenodo, \dodoi{10.5281/zenodo.1419754}

\bibitem[{{Teague} \& {Loomis}(2020)}]{Teague2020}
{Teague}, R., \& {Loomis}, R. 2020, \apj, 899, 157,
  \dodoi{10.3847/1538-4357/aba956}

\bibitem[{{Teague} {et~al.}(2016){Teague}, {Guilloteau}, {Semenov}, {Henning},
  {Dutrey}, {Pi{\'e}tu}, {Birnstiel}, {Chapillon}, {Hollenbach}, \&
  {Gorti}}]{Teague2016}
{Teague}, R., {Guilloteau}, S., {Semenov}, D., {et~al.} 2016, \aap, 592, A49,
  \dodoi{10.1051/0004-6361/201628550}

\bibitem[{{Thi} {et~al.}(2004){Thi}, {van Zadelhoff}, \& {van
  Dishoeck}}]{Thi2004}
{Thi}, W.~F., {van Zadelhoff}, G.~J., \& {van Dishoeck}, E.~F. 2004, \aap, 425,
  955, \dodoi{10.1051/0004-6361:200400026}

\bibitem[{{Tielens} \& {Hollenbach}(1985)}]{Tielens1985}
{Tielens}, A.~G.~G.~M., \& {Hollenbach}, D. 1985, \apj, 291, 722,
  \dodoi{10.1086/163111}

\bibitem[{{van der Walt} {et~al.}(2011){van der Walt}, {Colbert}, \&
  {Varoquaux}}]{vanderWalt2011}
{van der Walt}, S., {Colbert}, S.~C., \& {Varoquaux}, G. 2011, Computing in
  Science and Engineering, 13, 22, \dodoi{10.1109/MCSE.2011.37}

\bibitem[{{van Terwisga} {et~al.}(2019){van Terwisga}, {van Dishoeck},
  {Cazzoletti}, {Facchini}, {Trapman}, {Williams}, {Manara}, {Miotello}, {van
  der Marel}, {Ansdell}, {Hogerheijde}, {Tazzari}, \&
  {Testi}}]{vanTerwisga2019}
{van Terwisga}, S.~E., {van Dishoeck}, E.~F., {Cazzoletti}, P., {et~al.} 2019,
  \aap, 623, A150, \dodoi{10.1051/0004-6361/201834257}

\bibitem[{{van Zadelhoff} {et~al.}(2003){van Zadelhoff}, {Aikawa},
  {Hogerheijde}, \& {van Dishoeck}}]{vanZadelhoff2003}
{van Zadelhoff}, G.~J., {Aikawa}, Y., {Hogerheijde}, M.~R., \& {van Dishoeck},
  E.~F. 2003, \aap, 397, 789, \dodoi{10.1051/0004-6361:20021592}

\bibitem[{{Virtanen} {et~al.}(2020){Virtanen}, {Gommers}, {Oliphant},
  {Haberland}, {Reddy}, {Cournapeau}, {Burovski}, {Peterson}, {Weckesser},
  {Bright}, {van der Walt}, {Brett}, {Wilson}, {Jarrod Millman}, {Mayorov},
  {Nelson}, {Jones}, {Kern}, {Larson}, {Carey}, {Polat}, {Feng}, {Moore}, {Vand
  erPlas}, {Laxalde}, {Perktold}, {Cimrman}, {Henriksen}, {Quintero}, {Harris},
  {Archibald}, {Ribeiro}, {Pedregosa}, {van Mulbregt}, \&
  {Contributors}}]{SciPy2020}
{Virtanen}, P., {Gommers}, R., {Oliphant}, T.~E., {et~al.} 2020, Nature
  Methods, 17, 261, \dodoi{https://doi.org/10.1038/s41592-019-0686-2}

\bibitem[{{Visser} {et~al.}(2018){Visser}, {Bruderer}, {Cazzoletti},
  {Facchini}, {Heays}, \& {van Dishoeck}}]{Visser2018}
{Visser}, R., {Bruderer}, S., {Cazzoletti}, P., {et~al.} 2018, \aap, 615, A75,
  \dodoi{10.1051/0004-6361/201731898}

\bibitem[{{Yen} {et~al.}(2016){Yen}, {Koch}, {Liu}, {Puspitaningrum}, {Hirano},
  {Lee}, \& {Takakuwa}}]{Yen2016}
{Yen}, H.-W., {Koch}, P.~M., {Liu}, H.~B., {et~al.} 2016, \apj, 832, 204,
  \dodoi{10.3847/0004-637X/832/2/204}

\bibitem[{{Zhang} {et~al.}(2021){Zhang}, {Booth}, {Law}, {Bosman}, {Schwarz},
  {Bergin}, {{\"O}berg}, {Andrews}, {Guzm{\'a}n}, {Walsh}, {Qi}, {van 't Hoff},
  {Long}, {Wilner}, {Huang}, {Czekala}, {Ilee}, {Cataldi}, {Bergner}, {Aikawa},
  {Teague}, {Bae}, {Loomis}, {Calahan}, {Alarc{\'o}n}, {M{\'e}nard}, {Le Gal},
  {Sierra}, {Yamato}, {Nomura}, {Tsukagoshi}, {P{\'e}rez}, {Trapman}, {Liu}, \&
  {Furuya}}]{Zhang2020}
{Zhang}, K., {Booth}, A.~S., {Law}, C.~J., {et~al.} 2021, arXiv e-prints,
  arXiv:2109.06233.
\newblock \doarXiv{2109.06233}

\end{thebibliography}

\end{document}